\documentclass[aps,prb,floatfix,twocolumn,superscriptaddress,showpacs,amsmath,amssymb]{revtex4-1}

\usepackage{graphicx}
\usepackage{epsfig} 
\usepackage{amssymb}
\usepackage{amsmath}
\usepackage{color}
\usepackage{tabularx}
\usepackage{url}
\usepackage{hyperref}
\usepackage{capt-of}

\newcolumntype{C}[1]{>{\centering\let\newline\\\arraybackslash\hspace{0pt}}m{#1}}

\allowdisplaybreaks

\begin{document}

\title{Screening in orbital-density-dependent functionals} 

\author{Nicola Colonna}
\affiliation{Theory and Simulation of Materials (THEOS) \\ and National Centre for Computational Design and Discovery of Novel Materials (MARVEL),
\'Ecole Polytechnique F\' ed\'erale de Lausanne, 1015 Lausanne, Switzerland} 
\author{Ngoc Linh Nguyen}
\affiliation{Theory and Simulation of Materials (THEOS) \\ and National Centre for Computational Design and Discovery of Novel Materials (MARVEL),
\'Ecole Polytechnique F\' ed\'erale de Lausanne, 1015 Lausanne, Switzerland}
\author{Andrea Ferretti}
\affiliation{Centro S3, CNR-Istituto Nanoscienze, 41125 Modena, Italy} 
\author{Nicola Marzari}
\affiliation{Theory and Simulation of Materials (THEOS) \\ and National Centre for Computational Design and Discovery of Novel Materials (MARVEL),
\'Ecole Polytechnique F\' ed\'erale de Lausanne, 1015 Lausanne, Switzerland}

\date{\today}

\begin{abstract}
Electronic-structure functionals that include screening effects,
such as Hubbard or Koopmans' functionals, require to describe 
the response of a system to the fractional addition or removal of an electron
from an orbital or a manifold. Here, we present a general method to
incorporate screening based on linear-response theory, and we apply
it to the case of the orbital-by-orbital screening of Koopmans' functionals.
We illustrate the importance of such generalization when dealing with
challenging systems containing orbitals with very different chemical
character, also highlighting the simple dependence of the screening on the localization
of the  orbitals. We choose a set of 46 transition-metal complexes for which 
experimental data and accurate many-body perturbation theory calculations
are available. When compared to experiment, results for ionization potentials show a very
good performance with a mean absolute error of $~0.2$ eV, comparable to the most
accurate many-body perturbation theory approaches. These results reiterate the 
role of Koopmans' compliant functionals as simple and accurate quasiparticle
approximations to the exact spectral functional, bypassing diagrammatic expansions
and relying only on the physics of the local density or generalized-gradient approximation.
\end{abstract}

\maketitle

\section{Introduction} 

Accurate prediction of ground- and excited state properties of molecules can be made using quantum chemistry wave-function method
or many-body perturbation theory (MBPT) techniques~\cite{onida_electronic_2002}. However such calculations scale unfavorably with
the size of the system and become soon computationally untreatable.
For this reason electronic-structure approaches such as Hartree Fock (HF) or Kohn-Sham density-functional theory (KS-DFT) 
have often been used as a proxy to classify and understand excitation spectra. However, the eigenvalues of the KS potential
have no obvious relationship with the real excited states of the system. One notable exception is the first ionization potential
in finite systems, that is exactly reproduced in exact KS-DFT~\cite{perdew_density-functional_1982, perdew_physical_1983,almbladh_exact_1985,perdew_comment_1997}, 
but usually severely underestimated by standard local or semilocal approximations to the exchange-correlation energy functional.
In HF theory, the single-particle energies do have the physical meaning of excitation energies thanks to Koopmans' theorem, 
but miss important relaxation effects related to the addition of an electron (or a hole) to the system. These effects can be 
included in finite systems using e.g. $\Delta$ self consistent field  ($\Delta$SCF) calculations~\cite{onida_electronic_2002}, 
where the change in the energy associated with an electron addition/removal is calculated via two self-consistent calculations
done with N and N$+1$/N$-1$ electrons. However, it is not straightforward to apply this approach to single-particle energies 
beyond the frontier ones and its extension to solids poses some issues since the $\Delta$SCF correction, computed with 
standard density functional approximations, vanishes in the thermodynamic limit.~\cite{godby_density-relaxation_1998, perdew_understanding_2017}

For these reasons it would be highly desirable to have a functional yielding accurate single particle energies in addition to 
the well-established accuracy for ground state properties. Failures of standard density functional approximations 
in reproducing spectral quantities, such as ionization potentials and electron affinities, has been
connected to the deviation from piecewise linearity (PWL) of the the total energy functional as a function of particle number, 
and the associated lack of derivative discontinuity at integer particle numbers. First, the deviation from PWL
has been suggested~\cite{cococcioni_linear_2005, cohen_insights_2008, kulik_density_2006, mori-sanchez_many-electron_2006,mori-sanchez_localization_2008}
as a definition of electronic self-interaction errors~\cite{perdew_self-interaction_1981}, and in recently developed functionals,
such as DFT-corrected~\cite{zheng_improving_2011,kraisler_piecewise_2013,gorling_exchange-correlation_2015,li_local_2015},
range-separated~\cite{stein_fundamental_2010, kronik_excitation_2012, refaely-abramson_quasiparticle_2012} or dielectric-dependent 
hybrid functionals~\cite{shimazaki_band_2008, skone_self-consistent_2014,brawand_generalization_2016}, PWL has been recognized as
a critical feature to address. 
The criterion of PWL was in particular chosen as a key feature by some of us to introduce the class of Koopmans-compliant (KC) functionals~\cite{dabo_non-koopmans_2009,dabo_koopmans_2010,dabo_piecewise_2013,dabo_donor_2012,borghi_koopmans-compliant_2014}.
When used to purify approximate standard local or semilocal density functionals, Koopmans' corrections lead to orbital-density
dependent potentials that can be interpreted~\cite{ferretti_bridging_2014} as an approximation of the spectral potential, 
i.e. the local and dynamical potential that is necessary and sufficient to describe the local spectral density and, in turn, photoemission spectra~\cite{gatti_transforming_2007}.
With a relatively small increase of the computational cost this class of functionals delivers accurate spectral properties retaining
(in some cases exactly) the good performance of the underlying DFT functional for the ground state total energy~\cite{borghi_koopmans-compliant_2014}. 
The performance of the KC predictions greatly depends on the correct description of screening/relaxation effects associated 
to particle number modifications. For simple systems, these effects can be effectively captured by introducing a screening 
coefficient derived imposing the generalized Koopmans theorem on the frontier orbital~\cite{dabo_koopmans_2010, borghi_koopmans-compliant_2014}.
However, systems with a diverse electronic manifold, like for instance transition-metal compounds, call for a more 
accurate treatment since the same coefficient cannot equally well describe screening when the electron is removed (or added) 
e.g. from an $s$- or a $d$-like orbital. In this work we discuss how to correctly and fully include screening via 
orbital-dependent coefficients obtained from a linear response theory and thus ultimately from the dielectric screening. 
We then analyze the performance of the Koopmans-Integral (KI) functional~\cite{borghi_koopmans-compliant_2014}, a 
standard flavour of the KC class whose definition will be reviewed in Sec.~\ref{sec:method}, for a set of 46 
transition-metal complexes, and validate the results against GW calculations and experiments. 

The paper is organized as follow. In section~\ref{sec:method} we review the basic feature of Koopmans-compliant functionals 
and we introduce the linear-response approach to screening and relaxation. In section~\ref{sec.results} we compare ionization 
potentials for the set of molecules under study, obtained at different level of theory, with experimental data and 
state-of-the-art many-body perturbation theory calculations, followed by summary and conclusions.

\section{Method and computational details}
\label{sec:method}

In this section the main features of the KC class of functionals are reviewed with particular emphasis on the KI flavour,
and the new scheme to capture screening and relaxation effects based on a linear-response approach is introduced and discussed.

\subsection{Koopmans-compliant functionals}

Koopmans-compliant functionals~\cite{dabo_non-koopmans_2009,dabo_koopmans_2010,dabo_piecewise_2013,dabo_donor_2012,borghi_koopmans-compliant_2014}
explicitly enforce a generalized criterion of piecewise linearity (PWL) with respect to the fractional removal/addition
of an electron from any orbital in approximate DFT functionals. 
This is done by removing orbital-by-orbital the Slater contribution to the total energy, an approximately quadratic term
in the orbital occupation $f_i$, and replacing it with a linear (Koopmans) term. In practice, the Koopmans
correction is made up by the product of two equally important terms: ($i$) an orbital-density dependent corrections $\Pi^{\rm u}_i$
aiming at correctly describing the addition/removal of an electron in a {\it frozen-orbital} or  {\it unrelaxed} picture 
(restricted Koopmans theorem), and ($ii$) a screening factor $\alpha_i$ which takes into account the relaxation of the orbitals
as a response to the addition/removal process. In what follows we restrict our attention to the KI energy 
functional~\cite{borghi_koopmans-compliant_2014}, defined as $E^{\rm KI}=E^{\rm DFT}+\sum_i \alpha_i \Pi^{\rm uKI}_i$, where 
\begin{widetext}
 \begin{align}
 \Pi^{\rm uKI}_i(f_i) = & -\int_0^{f_i}ds \, \langle \psi_i | H^{\rm DFT}(s)| \psi_i \rangle \nonumber + f_i \int_0^1 ds \, \langle \psi_i |H^{\rm DFT}(s)|\psi_i\rangle  \nonumber \\
                         = & - \left\{ E^{\rm DFT}[\rho_u] - E^{\rm DFT}[\rho_u^{f_i=0}] \right\} + f_i \left\{ E^{\rm DFT}[\rho_u^{f_i=1}] -E^{\rm DFT}[\rho_u^{f_i=0}] \right\}
 \label{eq.uPiKi}
\end{align}
\end{widetext}
is the {\it unrelaxed} KI correction to the DFT energy (see Ref.~\citenum{borghi_koopmans-compliant_2014} for a 
detailed discussion of the KI energy functional and KI orbital-dependent potentials) and $\alpha_i$ the orbital
dependent screening factor. Here the orbitals $\{\psi_i\}$ are kept frozen, $\rho_u=\sum_j f_j |\psi_j|^2$ is 
the total density of the system and $H^{\rm DFT}(s)$ is the approximate KS-DFT Hamiltonian calculated at the 
unrelaxed density $\rho_u^{f_i=s}=\sum_{j\neq i} f_j |\psi_j|^2+ s|\psi_i|^2$, where only the explicit dependence
on the occupation $s$ is considered [the $f_j$ for $j \ne i$ are typically 1 (or 2 for spin degeneracy)].

At variance with DFT, the variation of the Koopmans-compliant functionals (and ODD functionals in general)
leads to local but orbital-dependent Hamiltonians that are in general not invariant under unitary transformations
of the electronic wavefunctions. The energy minimization defines a unique set of {\it variational} orbitals
that are usually very localized~\cite{heaton_self-interaction_1983, pederson_localdensity_1984, pederson_densityfunctional_1985, pederson_localized_1988, borghi_koopmans-compliant_2014, lehtola_variational_2014}
in space and resemble Boys orbitals~\cite{boys_construction_1960} or maximally localized Wannier functions~\cite{marzari_maximally_2012}.
At the minimum, the matrix of Lagrangian multipliers $\Lambda$, associated to the orthogonality constraint,
becomes Hermitian~\cite{goedecker_critical_1997, stengel_self-interaction_2008, borghi_variational_2015}
and can be diagonalized via a unitary transformation, allowing~\cite{pederson_localdensity_1984,korzdorfer_self-interaction_2008}
one to define a second set of {\it canonical} orbitals, usually less localized than the variational ones.
Although still a debated point~\cite{vydrov_tests_2007}, it is a common practice to interpret the eigenvalues
of $\Lambda$ as (canonical) orbital energies~\cite{pederson_localdensity_1984, vydrov_ionization_2005, korzdorfer_self-interaction_2008},
as clearly argued for in Ref.~\citenum{stengel_self-interaction_2008} and~\citenum{ferretti_bridging_2014}.

We note that the KI energy functional at integer occupation numbers preserves the unitary invariance
of the underlying DFT functional~\cite{borghi_koopmans-compliant_2014}. The energy minimization 
is therefore not sufficient to uniquely determine the variational orbitals; at the same time the 
KI potentials defining the KI eigenvalues depend on the actual representation of the electronic 
manifold~\cite{borghi_koopmans-compliant_2014}.
In previous works~\cite{borghi_koopmans-compliant_2014,nguyen_first-principles_2015,nguyen_first-principles_2016}
we removed this ambiguity by adding to the KI energy functional a vanishingly small Perdew-Zunger (PZ)
self-interaction correction, thus interpreting KI as the limit of the KIPZ functional~\cite{borghi_koopmans-compliant_2014}
when the PZ correction goes to zero.
The small PZ correction breaks the unitary invariance, and leads, through the energy functional 
minimization, to a set of well defined and typically localized variational orbitals (PZs orbitals
in the following) on which the KI corrections are computed. This choice is of course arbitrary,
although having a localized set of the orbitals is a key enabler for the good performance 
of KC functionals, and a mandatory one when extended systems are considered~\cite{nguyen_koopmans-compliant_2017}.
This is confirmed also here, by applying the KI corrections on different sets of orbitals 
with different degrees of localization; beside the set defined by the limiting procedure 
described above (PZs) we also consider the one represented by the canonical (KS) orbitals
of the base DFT functional (usually delocalized over the whole molecule) and other two 
localized representations given by maximally localized Wannier functions (MLWFs) and 
by atomic-projected Wannier functions (ProjWFs). The latter are obtained by a projection
of the KS states on a set of atomic orbitals that serves as a rough guess for the Wannier
function, followed by a symmetric orthonormalization. For the MLWFs, the localization is
enforced by using the sum of the quadratic spreads of the wavefunctions as a localization
criterion, and searching for the optimal unitary transformation that satisfies that 
criterion.~\cite{marzari_maximally_2012} The use of Wannier functions as a representation 
to apply the Koopmans' corrections and restore the missing piecewise linearity of approximate
DFT functionals has been shown to give good results for the band gaps of solids~\cite{ma_using_2016}.
In the following we use the notation KI@[representation] to indicate the set of orbitals used
to compute the KI corrections.

The screening and relaxation effects, naturally associated with the removal/addition of an 
electron, are accounted for by the multiplicative factor $\alpha_i$ in front of the bare 
correction $\Pi^{\rm uKI}_i$. How to determine this screening is the central goal of this work. 
In previous applications we showed that a unique (identical for all valence orbitals) screening 
factor, chosen to enforce the generalized Koopmans' condition on the frontier orbital~\cite{dabo_koopmans_2010,borghi_koopmans-compliant_2014},
is sufficient to accurately predict the ionization potentials (IPs) and also photoemission
spectra in a variety of molecular systems~\cite{borghi_koopmans-compliant_2014, nguyen_first-principles_2015, nguyen_first-principles_2016}. 
A second screening coefficient, derived imposing the Koopmans' condition on the lowest unoccupied 
moleclar orbital (LUMO), can be attached to all the conduction states, extending the 
predictive power of KC functionals to electron affinities (EAs)~\cite{nguyen_first-principles_2015, nguyen_first-principles_2016}. 
However, a more sophisticated orbital-dependent choice is needed in case of systems
with a more diverse electronic manifold. In the next section we present and discuss this more general,
physically sound and orbital-dependent treatment of the screening based on linear-response theory.~\cite{martin_interacting_2016,gross_local_1985}

\subsection{Screening in KC functionals}
In order to revert the unrelaxed Koopmans' correction to
a fully relaxed one, the screening coefficients can be formally defined as $\alpha_i = \Pi^{\rm rKI}_i / \Pi^{\rm uKI}_i$ where $\Pi_i^{\rm rKI}$ is the 
{\it relaxed} Koopmans correction defined in Ref.~\citenum{dabo_koopmans_2010}. The latter can be  written as
\begin{widetext}
\begin{align}
 \Pi^{\rm rKI}_i(f_i) = & - \left\{ E^{\rm DFT}[\rho] - E^{\rm DFT}[\rho^{f_i=0}] \right\}
                         + f_i \left\{ E^{\rm DFT}[\rho^{f_i=1}] -E^{\rm DFT}[\rho^{f_i=0}] \right\},
\label{eq.rPiKi}
\end{align}
\end{widetext}
where, at variance with Eq.~(\ref{eq.uPiKi}), $\rho^{f_i=s}$ in this expression corresponds to the fully relaxed 
density compatible with the condition $f_i=s$. Total energies with the additional constraint on one 
of the occupation numbers appearing in Eq.~(\ref{eq.rPiKi}), could in principle be evaluated by means of
constrained-density-functional approach~\cite{dederichs_ground_1984}.
Applications have been made to e.g. coulomb-interaction parameters to be used in model 
Hamiltonian~\cite{anisimov_density-functional_1991, hybertsen_calculation_1989, mcmahan_calculated_1988} or 
in the context of Hubbard $U$ correction to DFT~\cite{cococcioni_linear_2005}. Here we follow an alternative route, and
evaluate each term in Eq.~(\ref{eq.rPiKi}) in a perturbative way introducing a Taylor expansion of the DFT 
energy with respect to the occupation $f_i$ around some reference occupation $f_{\rm ref}$:
\begin{equation}
 E^{\rm DFT}[\rho^{f_i=s}] =  \sum_n \frac{1}{n!} \left. \frac{d^n E^{\rm DFT}}{d f_i^n}\right|_{f_{ref}}(s-f_{\rm ref})^n.
\end{equation}
Substituting in Eq.~(\ref{eq.rPiKi}) and stopping at the second order we find:
\begin{equation}
 \Pi^{\rm rKI}_i(f_i) = \frac{1}{2}f_i(1-f_i)\left. \frac{d^2 E^{\rm DFT}}{d f_i^2}\right|_{f_{\rm ref}} + O(f_i^3).
 \label{eq.2rPiKi}
\end{equation}
Due to Janak's theorem~\cite{janak_proof_1978} the second derivative of the energy wrt a given occupation represents also the first derivative of the corresponding eigenvalue.
In the following, we will work in the diagonal representation of the KS-DFT 
Hamiltonian\footnote{This restriction can be however released (see Supporting Information) and the final result of the derivation 
[Eq.~(\ref{eq.2urPiKi})] apply to any equivalent representation of the KS orbitals, i.e. to any set of orbitals related to the KS one 
by a unitary transformation.} and in the general case of relaxed orbitals, the frozen-orbital case being recovered trivially at the end of the derivation. We have:
\begin{align}
\frac{d^2 E^{\rm DFT}}{df_i^2} & = \frac{d \varepsilon_i}{df_i} = \langle \psi_i | \frac{dv_{\rm Hxc}}{df_i} | \psi_i \rangle  \nonumber \\
                                & = \int d\mathbf{r} d\mathbf{r}' n_i(\mathbf{r}) f_{\rm Hxc}[\rho](\mathbf{r}, \mathbf{r}') \frac{d\rho(\mathbf{r}')}{d f_i}
\label{eq.d2E_d2fi}                                
\end{align}
where we have used the Hellmann-Feynman theorem~\cite{hellmann_einfuhrung_1944, feynman_forces_1939} in the second identity, and
have introduced the orbital density $n_i = |\psi_i|^2$ and the Hartree-exchange-correlation (Hxc) kernel $f_{\rm Hxc}=\delta^2 
E_{\rm Hxc}/\delta \rho^2$~\cite{petersilka_excitation_1996,nalewajski_density_1996}. 
The derivative of the charge density with respect to the occupation is made up by two contributions: the first one comes from the 
explicit dependence of the density on the occupations while the second one comes from the change in the orbitals at fixed 
occupation, i.e. at fixed number of particles:
\begin{align}
 \frac{d\rho(\mathbf{r})}{df_i} = & n_i(\mathbf{r}) + \int d\mathbf{r}' \frac{\delta \rho(\mathbf{r})}{\delta v_{\rm KS}(\mathbf{r}')} \frac{d v_{\rm KS}(\mathbf{r}')}{df_i}
 \nonumber \\
  = & n_i(\mathbf{r}) + \int d\mathbf{r}' \chi_0(\mathbf{r},\mathbf{r}') \frac{d v_{\rm Hxc}(\mathbf{r}')}{df_i} 
 \nonumber\\
  = & n_i(\mathbf{r}) + \int d\mathbf{r}' \left[\chi_0 f_{\rm Hxc}\right]_{(\mathbf{r},\mathbf{r}')} \frac{d\rho(\mathbf{r}')}{df_i}
  \label{eq.drho_dfi}
\end{align}
where $\chi_0=\delta \rho/ \delta v_{\rm KS}$ is the KS (non-interacting) density-density response function. 
Equation~(\ref{eq.drho_dfi}) is a Dyson-like equation for the derivative of the charge density. Its iterative solution can be recast in a compact
form introducing the interacting density-density response function $\chi=\chi_0 + \chi_0 f_{\rm Hxc} \chi$~\cite{gross_local_1985,petersilka_excitation_1996}:
\begin{equation}
 \frac{d\rho(\mathbf{r})}{df_i} = n_i(\mathbf{r}) + \int d\mathbf{r}' \left[\chi f_{\rm Hxc}\right]_{(\mathbf{r},\mathbf{r}')} n_i(\mathbf{r}').
 \label{eq.drho_dfi_2}
\end{equation}
where it is understood that the response function and the Hxc-kernel are evaluated at $\rho=\rho^{f_i =f_{\rm ref}}$. 
In the frozen orbital approximation the second term on the right hand side of the equation above is exactly zero.
Combining Eqs.~(\ref{eq.drho_dfi_2}),~(\ref{eq.d2E_d2fi}) and~(\ref{eq.2rPiKi}) we obtain the central result of this paper:
\begin{align}
 \Pi^{\rm (2)uKI}_i(f_i) & = \frac{1}{2}f_i(1-f_i)\int d\mathbf{r} d\mathbf{r}' n_i(\mathbf{r})f_{\rm Hxc}(\mathbf{r},\mathbf{r}')n_i(\mathbf{r}') \nonumber \\
 \Pi^{\rm (2)rKI}_i(f_i) & = \frac{1}{2}f_i(1-f_i)\int d\mathbf{r} d\mathbf{r}' n_i(\mathbf{r}) \mathcal{F}_{\rm Hxc}(\mathbf{r},\mathbf{r}')n_i(\mathbf{r}') 
 \label{eq.2urPiKi}
\end{align}
where we have defined the screened Hxc-kernel $ \mathcal{F}_{\rm Hxc} = (I+f_{\rm Hxc}\chi )f_{\rm Hxc} = \epsilon^{-1}f_{\rm Hxc}$. 
It is important to stress that all the quantities needed to evaluate Eqs.~(\ref{eq.2urPiKi}),
i.e. the orbital density $n_i(\mathbf{r})$, the Hxc kernel $f_{\rm Hxc}$ and the response functions $\chi_0$ and $\chi$, are
all ground-state properties and, therefore, accessible from the reference ground state calculation
(usually the one for the neutral system). 
Instead, in a finite difference approach, as the one adopted in Refs.~\citenum{nguyen_koopmans-compliant_2017, ma_using_2016}
to compute energy differences when changing the occupation numbers, one needs to ask for the additional requirement
to keep fixed the orbital where the electron is added or removed. This is to prevent it from morphing into the highest (partially)
occupied orbital, as this would always be the most favorable energetic configuration because of the Aufbau principle. 
In the linear response approach described above this is not needed since one always refers to single-particle
orbitals and energies of the reference calculation, which are fixed by construction.

Even if derived from a simplified treatment based on a second order Taylor expansion, it is formally evident that the inclusion of orbital 
relaxation leads to a screening of the unrelaxed Koopmans correction. The integrals appearing in Eq.~(\ref{eq.2urPiKi})
can be interpreted as the effective interaction between the electrons in the orbital $i$ when all the other orbitals are allowed (second line) 
or not (first line) to readjust. 
The connection between the second derivative of the energy with respect to the occupation of a localized orbital and the effective interaction
between localized electrons has been also discussed in the context of Hubbard corrections to DFT~\cite{anisimov_band_1991,anisimov_density-functional_1993,liechtenstein_density-functional_1995}. The value of the static $U$ parameter to be used in 
the model Hamiltonian is indeed determined from the constrained variation of the DFT eigenvalue with respect to the occupation
number of the localized orbitals~\cite{mcmahan_calculated_1988, hybertsen_calculation_1989, gunnarsson_density-functional_1989, gunnarsson_calculation_1990, anisimov_density-functional_1991, cococcioni_linear_2005}. The frequency-dependence of
the effective interaction has been also computed within the constrained random-phase approximation~\cite{springer_frequency-dependent_1998, kotani_ab_2000,aryasetiawan_frequency-dependent_2004,aryasetiawan_calculations_2006} (cRPA)
using an expression similar to the second line of Eq.~(\ref{eq.2urPiKi}) but evaluated at the RPA level, i.e. neglecting the xc-kernel 
both in the Dyson-like equation defining the response function $\chi$ and in the dielectric matrix $\epsilon$.
Within this second order expansion the similarity of the Koopmans-compliant functional with the $+U$ correction in DFT is even more evident and indeed these functionals can be interpreted as a generalization of the DFT+U approach to the entire electronic manifold (this was the reason, in primis, of their introduction~\cite{dabo_non-koopmans_2009,dabo_koopmans_2010}). 
Notwithstanding the apparent similarity, there is a fundamental 
difference between the two approaches in the fact that KC corrections to DFT aim at describing addition or removal of an electron
from the system (charged excitation), while the $+U$ correction can be interpreted as the energy cost associated to move a fraction 
of an electron from a localized orbital (or manifold) to the bath represented by the rest of the system (neutral excitation). 
Then, as argued in Ref.~\citenum{solovyev_screening_2005}, the renormalization of the bare interaction might take place through
different screening channels depending on whether the electron is added/removed or continues to stay in the system. 
In the derivation above, the explicit variation of the particle number is considered [first contribution in
Eq.~(\ref{eq.drho_dfi_2})], as KC functionals aim at describing charged excitations, 
and the screened Koopmans' correction in Eq.~(\ref{eq.2urPiKi}) is thus the correct one for such processes.
Most importantly, the central results of Eq.~(\ref{eq.2urPiKi}) are still valid if 
one substitutes the canonical set of KS orbitals with another set related to the first one by a unitary transformation
(see Supporting Information); this is extremely important because the screening coefficients to be used in KC functionals are actually connected to the 
variational orbitals, i.e. the orbitals that minimize the KC functionals, and these are related by a unitary transformation
to the canonical orbitals. 

The second line in Eq.~(\ref{eq.2urPiKi}) could be used as a relaxed Koopmans correction, albeit exact only up to second order. 
We therefore continue to use the original definition of the screening, and we introduce 
an orbital-dependent screening coefficient $\alpha_i$ defined as the ratio between the relaxed and unrelaxed 
second-order Koopmans correction:
\begin{equation}
 \alpha_i = \frac{ \Pi^{\rm (2)rKI}_i(f_i)}{\Pi^{\rm (2)uKI}_i(f_i)} = \frac{\langle n_i | \mathcal{F}_{\rm Hxc} | n_i \rangle}{\langle n_i | f_{\rm Hxc} | n_i \rangle}
 \label{eq.alphai}
\end{equation}
with $\langle n_i | A | n_i \rangle = \int d\mathbf{r} d\mathbf{r}' \; n_i(\mathbf{r}) A(\mathbf{r},\mathbf{r}') n_i(\mathbf{r}')$. 
The simplified screening coefficient introduced in previous publications~\cite{dabo_koopmans_2010,borghi_koopmans-compliant_2014}
can be seen as a particular case of the present approach reducing the electronic screening function 
$\epsilon^{-1}(\mathbf{r},\mathbf{r}')$ to a constant.
We also note that the definition of the screening coefficient in Eq.~(\ref{eq.alphai}) is similar to the expression of the mixing parameter
in the context of dielectric-dependent hybrid functionals~\cite{skone_nonempirical_2016} generalized to finite systems~\cite{brawand_generalization_2016,brawand_performance_2017}. 

\begin{figure*}
\begin{minipage}{\textwidth} 
\captionof{figure}{Absolute difference in eV between calculated and experimental ionization potentials for different level of theory. 
The G$_0$W$_0$@PBE0 results are from Ref.~\citenum{korbel_benchmark_2014} (see Supporting information for a full table).}
\includegraphics[width=\textwidth]{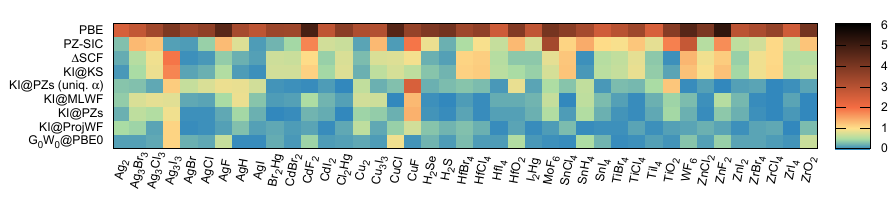}
\label{fig.color_map}
\vspace{0.5cm}
\noindent\rule{0.5\textwidth}{0.4pt}
 \centering
 \resizebox{\textwidth}{!}{%
 \begin{tabular}{l   C{1.2cm}           C{1.4cm}         C{1.4cm}         C{1.4cm}                 C{1.8cm}             C{2.1cm}           C{1.8cm}          C{2.3cm}         C{2.5cm} } 
   \hline\hline
   \noalign{\vskip 2mm}  
                              &  PBE            & PZ-SIC         & $\Delta$SCF      &  KI@KS          & KI@PZs (uniq.~${\alpha}$) & KI @MLWF           & KI@PZs           & KI @ProjWF    & G$_0$W$_0$ @PBE0  \\ 
     \noalign{\vskip 2.5mm}  
     \hline
     \noalign{\vskip 2.5mm}  
                   MAE        & 3.43            &  0.87           &  0.62            & 0.63            & 0.39                      & 0.29             & 0.22             & 0.20            & 0.21  \\
    \noalign{\vskip 2.5mm}
                   MSE        & -3.43           &  0.82           &  -0.51           & -0.57           & 0.28                      & 0.27             & 0.18             & -0.02           & -0.07 \\
     \noalign{\vskip 2.5mm}
                   Max($+$)   &  --             &  1.97 \hspace{1cm} (CuF)     &  0.96 \hspace{1cm} (CuF)      &  0.63 \hspace{1cm} (CuF)     &  2.51 \hspace{1cm} (CuF)               &  1.37 \hspace{1cm} (CuF)      &  1.43 \hspace{1cm} (CuF)      &  0.74 \hspace{1cm} (CuF)     & 0.53 \hspace{1cm} (SnH$_4$) \\
     \noalign{\vskip 2.5mm}  
                   Max($-$)   & 5.27 \hspace{1cm} (ZnF$_2$)  & 0.35 \hspace{1cm} (Ag$_2$)   & 1.25 \hspace{1cm} (SnCl$_4$)  & 1.30 \hspace{1cm} (ZnF$_2$ ) & 0.37 \hspace{1cm} (HfBr$_4$)           & 0.12 \hspace{1cm} (HfBr$_4$)  & 0.19 \hspace{1cm} (HfBr$_4$)  & 0.37 \hspace{1cm} (H$_2$Se)  & 0.47 \hspace{1cm} (CdF$_2$) \\
     \noalign{\vskip 2.5mm}
     \hline\hline
 \end{tabular}}
 \captionof{table}{Mean absolute error (MAE), mean signed error (MSE) and maximum (positive and negative)
 deviation between calculated and experimental ionization potentials for different level of theory (see
 text). For PBE all the errors are negative and only the maximum negative error makes sense and is reported. The G$_0$W$_0$@PBE0 calculations are from Ref.~\citenum{korbel_benchmark_2014}. All the data
 are in eV. The system corresponding to the maximum error is also reported in parenthesis.}
 \label{tab.MAE}
\end{minipage}
\end{figure*}

Screening coefficients have been efficiently computed resorting to the linear-response approach of density-functional perturbation theory (DFPT)~\cite{baroni_phonons_2001}.
The second term in the right hand side of Eq.~(\ref{eq.drho_dfi_2}) can be indeed interpreted as the density variation due to a ``bare'' perturbing potential 
$\Delta^i V(\mathbf{r}) = \Delta f_i \int d\mathbf{r}' \; f_{\rm Hxc}(\mathbf{r},\mathbf{r}') n_i(\mathbf{r}')$. 
This density variation is equivalent to that induced in the auxiliary KS system by an effective potential $\Delta^i V_{\text{eff}}(\mathbf{r}) = \Delta^i V(\mathbf{r}) +
\Delta^i V_{\rm Hxc}(\mathbf{r})$ with $\Delta^i V_{\rm Hxc}$ being the self consistent variation of the Hxc potential due to the change in the density. 
The linear-response calculation of the screening coefficient has been implemented in a modified version of the Phonon code of {\sc Quantum ESPRESSO}~\cite{giannozzi_quantum_2009,giannozzi_advanced_2017}.

\subsection{Computational setup}

We apply this complete orbital-dependent scheme to compute the IPs of a set of 46 transition metal complexes first introduced in Ref.~\citenum{korbel_benchmark_2014},
for which experimental as well as many body perturbation theory results are available (only 41 values are available from experiments).
The calculations are performed using Optimized Norm-Conserving 
Vanderbilt (ONCV) pseudopotentials~\cite{hamann_optimized_2013, schlipf_optimization_2015, ONCV_website} to model the interaction between nuclei and electrons.  
In order to simulate isolated molecules\footnote{The equilibrium geometries of the transition metal complexes were kindly provided by
the authors of Ref.~\citenum{korbel_benchmark_2014}.} we place these inside an orthorombic cell with 22 Bohr of vacuum in each direction, sufficient 
to converge total energies and single particle eigenvalues when the Coulomb interaction between periodic images are suppressed 
using reciprocal-space counter-charge corrections~\cite{li_electronic_2011}. 
The energy cut-off for the plane-wave expansion is set to 100 Ry (400 Ry for the charge density). 
All orbital-density dependent calculations presented here uses PBE~\cite{perdew_generalized_1996} as the underlying xc-energy functional; 
energy minimizations are performed on the space of complex wavefunctions.

The calculated IP is defined as minus the eigenvalue of the highest-occupied molecular orbital (HOMO) $ {\rm IP}=-\varepsilon_{ho}$ for all the theoretical methods reported, except for $\Delta$SCF when the IP is defined as the energy difference between the neutral molecule and its cation.

\section{Results and Discussion}~\label{sec.results}
All the calculated and available experimental IPs are listed in the Supporting Information. We focus below on the average performance of the KI functional as estimated by the mean absolute error (MAE), mean signed error (MSE) and maximum signed error [Max($+$), Max($-$)].

\subsection{Performance of the KI functional}
In Fig.~\ref{fig.color_map} the absolute difference between calculated and experimental ionization energies (IPs) is drawn as a color map. 
In Tab.~\ref{tab.MAE} the resulting mean absolute error (MAE), mean signed error (MSE) and maximum signed error [Max($+$), Max($-$)] are
listed\footnote{For those experiments for which is unclear whether they are vertical or adiabatic (see Supporting Information for the complete list) the 
calculated value are left out of the average. In the case of adiabatic experimental value we used the same correction used in Ref.~\citenum{korbel_benchmark_2014}
based on the difference between adiabatic and vertical $\Delta$SCF energies.}.
The PBE results show the well-known underestimation of the IPs due to the self-interaction error resulting in a MAE and a maximum signed error of 3.43  
and -5.27 eV, respectively. The self-interaction correction (SIC) by Perdew and Zunger~\cite{perdew_self-interaction_1981} (PZ-SIC in the Figures and Tables) over-corrects the
PBE results with an almost systematic overestimation of the IP, while $\Delta$SCF, that usually gives rather good estimation of the frontier orbitals energies,
in this case has a mean absolute error of 0.65 eV which is only slightly smaller than PZ-SIC. 
KI corrections computed on top of the canonical PBE orbitals (KI@KS) show an average performance that is comparable to that of $\Delta$SCF. 
The two methods perform very similarly for each single molecule (see Fig.~\ref{fig.correlation_map} and 
Tab.~I in Supporting Information); this is not by chance, and highlights the physical content embedded in the KI corrections:
When relaxation effects are correctly taken into account the KI functional reverts the KS eigenvalues into a $\Delta$SCF energy, inheriting the accuracy of finite-difference DFT energies.
The use of a localized representation leads to a significant improvement of the performance with a reduction of the MAE by a factor $2\div3$ over $\Delta$SCF
and KI@KS. The KI@ProjWF and the KI@PZs show very similar average performance with a MAE comparable to those from the best GW calculations, i.e. G$_0$W$_0$@PBE$0$
(MAE 0.21 eV), chosen among ten different prescriptions~\cite{korbel_benchmark_2014}.
We stress here that all the theoretical IPs do not include
relativistic nor zero-point motion effects. As highlighted in Refs.~\citenum{korbel_benchmark_2014, wu_photoelectron_2011}, the latter is usually negligible while the former might be more relevant for this set of molecules and affect the comparison between theory and experiment. Comparison between different theoretical methods is instead fully consistent since these all neglect relativistic effects and zero-point motion.

In order to highlight the importance of the orbital-dependent treatment of the screening we also show the KI@PZs results obtained using a unique value 
of the screening coefficient, i.e. $\alpha_i=\alpha_{cn} \; \; \forall i$ (uniq. $\alpha$ in plots and Tables). Here the subscript ``$cn$'' stands for 
``cation-neutral'' and points to the fact that $\alpha_{cn}$ is calculated~\cite{dabo_koopmans_2010,borghi_koopmans-compliant_2014,nguyen_first-principles_2016}
imposing the generalized Koopmans condition on the HOMO, i.e. requiring that the HOMO eigenvalue of the neutral system is equal to the LUMO eigenvalue of the same system deprived by one electron: $\rm{IP^{N}}(\alpha_{cn}) = \rm{EA^{(N-1)}(\alpha_{cn})}$.
A comparison between KI@PZs with and without the orbital-dependent treatment of the screening reveals that the MAE and the maximum errors are almost halved
highlighting the importance of a more advanced treatment of the screening for transition-metal complexes.
We also stress here that the way of computing a unique screening coefficient, although conceptually straightforward and appealing, is cumbersome to extend to 
orbitals different from the frontier ones, and also requires in practice multiple calculations at N and N-1 electrons~\cite{borghi_koopmans-compliant_2014}.
The linear-response approach introduced here bypasses both these problems in a natural way. 

For all our KI calculations, the outlier with the largest error with respect to experimental IP is the CuF molecule. It is highly unlikely that the large deviations
observed for this molecule (see Tab.~\ref{tab.MAE}) comes from effects not included in our calculations. Finite temperature and relativistic effects and/or zero-point motion could account for discrepancy of the order of few tenths of an electron volt~\cite{korbel_benchmark_2014}. However, it should be also mentioned that
the experiment values usually come with an error bar and in this particular case the experimentally measured ionization potential vary from 8.6 to 
10.9 eV~\cite{nist2005} making any comparison not particularly significant.

\subsection{Effect of the localized representation}

\begin{center}
\begin{figure}[t]
\includegraphics[width=8cm]{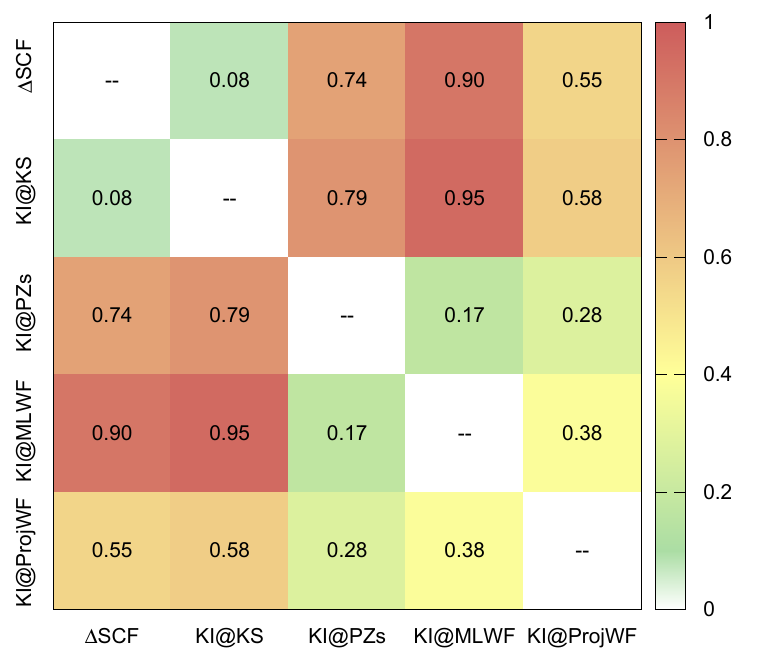}
\caption{Mean absolute distance (see text) between the results of the different KI calculations.}
\label{fig.correlation_map}
\end{figure} 
\end{center}

It's interesting to note that despite some inevitable dependence on the choice of the localized representation, there is a substantial equivalence 
between the KI calculations done on top of the 3 different localized set of orbitals. In Fig.~\ref{fig.correlation_map} we try to quantify such dependence 
of the IPs on the underlying representation defining the mean absolute distance (MAD) between two sets ($a$ and $b$) of IPs from different KI calculations as
\begin{equation}
 {\rm MAD}_{ab} = \frac{1}{N}\sum_{i=1}^N |{\rm IP}_i^a -{\rm IP}_i^b|,
\end{equation}
and plotting it as a color map. We also add the $\Delta$SCF results to highlight again its close relation with KI@KS. The MAD between these two sets of IPs is 
only 0.08 eV. From the correlation matrix one can also see that KI@PZs and KI@MLW are the second closest pair, reflecting the fact that the PZ localization
condition and the MLW one usually lead to very similar sets of orbitals. We also clearly see that calculations on localized sets of orbitals and 
calculations on KS states (KI@KS and $\Delta$SCF) form two distinct blocks with off-diagonal elements up to 1 eV. 
The better performance of KI when a localized representation is used with respect to the KI@KS emphasizes what is a key requirement for this class of functionals, i.e. expressing the Koopmans
corrections on a localized set of orbitals~\cite{nguyen_koopmans-compliant_2017}. While the discrepancy is not dramatic in the case of atoms or small molecules, it becomes more and more 
evident when increasing the size of the system; given the strong connection with the $\Delta$SCF method, KI@KS would experience the same 
failure of the finite-difference method in the thermodynamic limit~\cite{godby_density-relaxation_1998, perdew_understanding_2017}. Using a localized representation of the electronic manifold to compute the orbital
corrections ensures instead a finite correction also in the thermodynamic limit~\cite{nguyen_koopmans-compliant_2017}. In addition, there is a strong correlation between screening
parameters and orbital localization, e.g. as described by the
self-Hartree energies (see Supporting Information). Such correlations
would make it easier to apply this approach to large scale calculations,
where screening coefficients could be inferred from a few linear-response
tests.

Before concluding, we mention two points that are also relevant to the
present discussion and worthy of future studies. First, system
symmetries are not necessarily reproduced by orbital-density dependent
functionals~\cite{stengel_self-interaction_2008, lehtola_complex_2016}, if the corresponding Hamiltonians do not preserve these -
hence the symmetry of the localized representation can play an important
role, especially in small systems or in the atomic limit. Second, the
present discussion makes clear, as is already known for the case
of Hubbard functionals~\cite{kulik_accurate_2011,hsu_first-principles_2009}, that derivatives of the total energy
would need to take into account the dependence of the
screening coefficients on the varying parameters, e.g. atomic displacements.
This dependence is expected to be small or negligible, and in any case
these functionals are meant to great enhance the spectral properties,
leaving the energetics untouched or slight improved in strongly
self-interacting systems by the screened PZ correction; notably, for the
KI class of functionals the potential energy surface is
identical to the base functional, and thus such dependence is not relevant.

\section{CONCLUSIONS}

In conclusion we have shown how to extend the predictive power of Koopmans-compliant functionals to systems 
characterized by a complex electronic manifold, such as transition-metal complexes, finding that an orbital-dependent treatment 
of the screening is crucial to the correct description of spectral properties when orbitals with very different chemical 
character are considered. A scheme based on a linear-response approach has been introduced and discussed, highlighting 
the physical content behind the screening coefficient in Koopmans-compliant functionals. 
We found excellent agreement for the computed value of the ionization potentials with both experiment and state-of-the-art 
many-body perturbation theory, especially when the orbital dependence of the screening is correctly accounted for and when a localized representation
of the electronic manifold is used to enforce the generalized PWL condition of KC functionals. 
The results on a  carefully tested set of 41 transition-metal complexes show mean absolute errors of 0.20, 0.22 and 0.29 eV, respectively, 
for the KI functional calculated on different localized representations; ten different many-body perturbation theory approaches were in the range 
0.21 eV to 1.92 eV, with G$_0$W$_0$@PBE0 being the most accurate one.
These results reiterate the role of Koopmans' compliant functional
as spectral functionals able to simultaneously reproduce electronic
spectra and total energies, and as quasiparticle approximations to the
exact spectral potential~\cite{ferretti_bridging_2014, gatti_transforming_2007}.

\begin{acknowledgements}
We acknowledge partial support from the Swiss National
Centre for Computational Design and Discovery of Novel
Materials (MARVEL). NC acknowledges Dr. Matteo Coccoccioni 
for useful discussions. We would like to thank Dr. Sabine K\"orbel 
for providing the equilibrium geometries.
\end{acknowledgements}



\begin{thebibliography}{87}%
\makeatletter
\providecommand \@ifxundefined [1]{%
 \@ifx{#1\undefined}
}%
\providecommand \@ifnum [1]{%
 \ifnum #1\expandafter \@firstoftwo
 \else \expandafter \@secondoftwo
 \fi
}%
\providecommand \@ifx [1]{%
 \ifx #1\expandafter \@firstoftwo
 \else \expandafter \@secondoftwo
 \fi
}%
\providecommand \natexlab [1]{#1}%
\providecommand \enquote  [1]{``#1''}%
\providecommand \bibnamefont  [1]{#1}%
\providecommand \bibfnamefont [1]{#1}%
\providecommand \citenamefont [1]{#1}%
\providecommand \href@noop [0]{\@secondoftwo}%
\providecommand \href [0]{\begingroup \@sanitize@url \@href}%
\providecommand \@href[1]{\@@startlink{#1}\@@href}%
\providecommand \@@href[1]{\endgroup#1\@@endlink}%
\providecommand \@sanitize@url [0]{\catcode `\\12\catcode `\$12\catcode
  `\&12\catcode `\#12\catcode `\^12\catcode `\_12\catcode `\%12\relax}%
\providecommand \@@startlink[1]{}%
\providecommand \@@endlink[0]{}%
\providecommand \url  [0]{\begingroup\@sanitize@url \@url }%
\providecommand \@url [1]{\endgroup\@href {#1}{\urlprefix }}%
\providecommand \urlprefix  [0]{URL }%
\providecommand \Eprint [0]{\href }%
\providecommand \doibase [0]{http://dx.doi.org/}%
\providecommand \selectlanguage [0]{\@gobble}%
\providecommand \bibinfo  [0]{\@secondoftwo}%
\providecommand \bibfield  [0]{\@secondoftwo}%
\providecommand \translation [1]{[#1]}%
\providecommand \BibitemOpen [0]{}%
\providecommand \bibitemStop [0]{}%
\providecommand \bibitemNoStop [0]{.\EOS\space}%
\providecommand \EOS [0]{\spacefactor3000\relax}%
\providecommand \BibitemShut  [1]{\csname bibitem#1\endcsname}%
\let\auto@bib@innerbib\@empty
\bibitem [{\citenamefont {Onida}\ \emph {et~al.}(2002)\citenamefont {Onida},
  \citenamefont {Reining},\ and\ \citenamefont
  {Rubio}}]{onida_electronic_2002}%
  \BibitemOpen
  \bibfield  {author} {\bibinfo {author} {\bibfnamefont {G.}~\bibnamefont
  {Onida}}, \bibinfo {author} {\bibfnamefont {L.}~\bibnamefont {Reining}}, \
  and\ \bibinfo {author} {\bibfnamefont {A.}~\bibnamefont {Rubio}},\ }\href
  {\doibase 10.1103/RevModPhys.74.601} {\bibfield  {journal} {\bibinfo
  {journal} {Reviews of Modern Physics}\ }\textbf {\bibinfo {volume} {74}},\
  \bibinfo {pages} {601} (\bibinfo {year} {2002})}\BibitemShut {NoStop}%
\bibitem [{\citenamefont {Perdew}\ \emph {et~al.}(1982)\citenamefont {Perdew},
  \citenamefont {Parr}, \citenamefont {Levy},\ and\ \citenamefont
  {Balduz}}]{perdew_density-functional_1982}%
  \BibitemOpen
  \bibfield  {author} {\bibinfo {author} {\bibfnamefont {J.~P.}\ \bibnamefont
  {Perdew}}, \bibinfo {author} {\bibfnamefont {R.~G.}\ \bibnamefont {Parr}},
  \bibinfo {author} {\bibfnamefont {M.}~\bibnamefont {Levy}}, \ and\ \bibinfo
  {author} {\bibfnamefont {J.~L.}\ \bibnamefont {Balduz}},\ }\href {\doibase
  10.1103/PhysRevLetters49.1691} {\bibfield  {journal} {\bibinfo  {journal}
  {Physical Review Letters}\ }\textbf {\bibinfo {volume} {49}},\ \bibinfo
  {pages} {1691} (\bibinfo {year} {1982})}\BibitemShut {NoStop}%
\bibitem [{\citenamefont {Perdew}\ and\ \citenamefont
  {Levy}(1983)}]{perdew_physical_1983}%
  \BibitemOpen
  \bibfield  {author} {\bibinfo {author} {\bibfnamefont {J.~P.}\ \bibnamefont
  {Perdew}}\ and\ \bibinfo {author} {\bibfnamefont {M.}~\bibnamefont {Levy}},\
  }\href {\doibase 10.1103/PhysRevLetters51.1884} {\bibfield  {journal}
  {\bibinfo  {journal} {Physical Review Letters}\ }\textbf {\bibinfo {volume}
  {51}},\ \bibinfo {pages} {1884} (\bibinfo {year} {1983})}\BibitemShut
  {NoStop}%
\bibitem [{\citenamefont {Almbladh}\ and\ \citenamefont {von
  Barth}(1985)}]{almbladh_exact_1985}%
  \BibitemOpen
  \bibfield  {author} {\bibinfo {author} {\bibfnamefont {C.-O.}\ \bibnamefont
  {Almbladh}}\ and\ \bibinfo {author} {\bibfnamefont {U.}~\bibnamefont {von
  Barth}},\ }\href {\doibase 10.1103/PhysRevB.31.3231} {\bibfield  {journal}
  {\bibinfo  {journal} {Physical Review B}\ }\textbf {\bibinfo {volume} {31}},\
  \bibinfo {pages} {3231} (\bibinfo {year} {1985})}\BibitemShut {NoStop}%
\bibitem [{\citenamefont {Perdew}\ and\ \citenamefont
  {Levy}(1997)}]{perdew_comment_1997}%
  \BibitemOpen
  \bibfield  {author} {\bibinfo {author} {\bibfnamefont {J.~P.}\ \bibnamefont
  {Perdew}}\ and\ \bibinfo {author} {\bibfnamefont {M.}~\bibnamefont {Levy}},\
  }\href {\doibase 10.1103/PhysRevB.56.16021} {\bibfield  {journal} {\bibinfo
  {journal} {Physical Review B}\ }\textbf {\bibinfo {volume} {56}},\ \bibinfo
  {pages} {16021} (\bibinfo {year} {1997})}\BibitemShut {NoStop}%
\bibitem [{\citenamefont {Godby}\ and\ \citenamefont
  {White}(1998)}]{godby_density-relaxation_1998}%
  \BibitemOpen
  \bibfield  {author} {\bibinfo {author} {\bibfnamefont {R.~W.}\ \bibnamefont
  {Godby}}\ and\ \bibinfo {author} {\bibfnamefont {I.~D.}\ \bibnamefont
  {White}},\ }\href {\doibase 10.1103/PhysRevLetters80.3161} {\bibfield
  {journal} {\bibinfo  {journal} {Physical Review Letters}\ }\textbf {\bibinfo
  {volume} {80}},\ \bibinfo {pages} {3161} (\bibinfo {year}
  {1998})}\BibitemShut {NoStop}%
\bibitem [{\citenamefont {Perdew}\ \emph {et~al.}(2017)\citenamefont {Perdew},
  \citenamefont {Yang}, \citenamefont {Burke}, \citenamefont {Yang},
  \citenamefont {Gross}, \citenamefont {Scheffler}, \citenamefont {Scuseria},
  \citenamefont {Henderson}, \citenamefont {Zhang}, \citenamefont {Ruzsinszky},
  \citenamefont {Peng}, \citenamefont {Sun}, \citenamefont {Trushin},\ and\
  \citenamefont {G{\"o}rling}}]{perdew_understanding_2017}%
  \BibitemOpen
  \bibfield  {author} {\bibinfo {author} {\bibfnamefont {J.~P.}\ \bibnamefont
  {Perdew}}, \bibinfo {author} {\bibfnamefont {W.}~\bibnamefont {Yang}},
  \bibinfo {author} {\bibfnamefont {K.}~\bibnamefont {Burke}}, \bibinfo
  {author} {\bibfnamefont {Z.}~\bibnamefont {Yang}}, \bibinfo {author}
  {\bibfnamefont {E.~K.~U.}\ \bibnamefont {Gross}}, \bibinfo {author}
  {\bibfnamefont {M.}~\bibnamefont {Scheffler}}, \bibinfo {author}
  {\bibfnamefont {G.~E.}\ \bibnamefont {Scuseria}}, \bibinfo {author}
  {\bibfnamefont {T.~M.}\ \bibnamefont {Henderson}}, \bibinfo {author}
  {\bibfnamefont {I.~Y.}\ \bibnamefont {Zhang}}, \bibinfo {author}
  {\bibfnamefont {A.}~\bibnamefont {Ruzsinszky}}, \bibinfo {author}
  {\bibfnamefont {H.}~\bibnamefont {Peng}}, \bibinfo {author} {\bibfnamefont
  {J.}~\bibnamefont {Sun}}, \bibinfo {author} {\bibfnamefont {E.}~\bibnamefont
  {Trushin}}, \ and\ \bibinfo {author} {\bibfnamefont {A.}~\bibnamefont
  {G{\"o}rling}},\ }\href {\doibase 10.1073/pnas.1621352114} {\bibfield
  {journal} {\bibinfo  {journal} {PNAS}\ }\textbf {\bibinfo {volume} {114}},\
  \bibinfo {pages} {2801} (\bibinfo {year} {2017})}\BibitemShut {NoStop}%
\bibitem [{\citenamefont {Cococcioni}\ and\ \citenamefont
  {de~Gironcoli}(2005)}]{cococcioni_linear_2005}%
  \BibitemOpen
  \bibfield  {author} {\bibinfo {author} {\bibfnamefont {M.}~\bibnamefont
  {Cococcioni}}\ and\ \bibinfo {author} {\bibfnamefont {S.}~\bibnamefont
  {de~Gironcoli}},\ }\href {\doibase 10.1103/PhysRevB.71.035105} {\bibfield
  {journal} {\bibinfo  {journal} {Physical Review B}\ }\textbf {\bibinfo
  {volume} {71}},\ \bibinfo {pages} {035105} (\bibinfo {year}
  {2005})}\BibitemShut {NoStop}%
\bibitem [{\citenamefont {Cohen}\ \emph {et~al.}(2008)\citenamefont {Cohen},
  \citenamefont {Mori-S{\'a}nchez},\ and\ \citenamefont
  {Yang}}]{cohen_insights_2008}%
  \BibitemOpen
  \bibfield  {author} {\bibinfo {author} {\bibfnamefont {A.~J.}\ \bibnamefont
  {Cohen}}, \bibinfo {author} {\bibfnamefont {P.}~\bibnamefont
  {Mori-S{\'a}nchez}}, \ and\ \bibinfo {author} {\bibfnamefont
  {W.}~\bibnamefont {Yang}},\ }\href {\doibase 10.1126/science.1158722}
  {\bibfield  {journal} {\bibinfo  {journal} {Science}\ }\textbf {\bibinfo
  {volume} {321}},\ \bibinfo {pages} {792} (\bibinfo {year}
  {2008})}\BibitemShut {NoStop}%
\bibitem [{\citenamefont {Kulik}\ \emph {et~al.}(2006)\citenamefont {Kulik},
  \citenamefont {Cococcioni}, \citenamefont {Scherlis},\ and\ \citenamefont
  {Marzari}}]{kulik_density_2006}%
  \BibitemOpen
  \bibfield  {author} {\bibinfo {author} {\bibfnamefont {H.~J.}\ \bibnamefont
  {Kulik}}, \bibinfo {author} {\bibfnamefont {M.}~\bibnamefont {Cococcioni}},
  \bibinfo {author} {\bibfnamefont {D.~A.}\ \bibnamefont {Scherlis}}, \ and\
  \bibinfo {author} {\bibfnamefont {N.}~\bibnamefont {Marzari}},\ }\href
  {\doibase 10.1103/PhysRevLetters97.103001} {\bibfield  {journal} {\bibinfo
  {journal} {Physical Review Letters}\ }\textbf {\bibinfo {volume} {97}},\
  \bibinfo {pages} {103001} (\bibinfo {year} {2006})}\BibitemShut {NoStop}%
\bibitem [{\citenamefont {Mori-S{\'a}nchez}\ \emph {et~al.}(2006)\citenamefont
  {Mori-S{\'a}nchez}, \citenamefont {Cohen},\ and\ \citenamefont
  {Yang}}]{mori-sanchez_many-electron_2006}%
  \BibitemOpen
  \bibfield  {author} {\bibinfo {author} {\bibfnamefont {P.}~\bibnamefont
  {Mori-S{\'a}nchez}}, \bibinfo {author} {\bibfnamefont {A.~J.}\ \bibnamefont
  {Cohen}}, \ and\ \bibinfo {author} {\bibfnamefont {W.}~\bibnamefont {Yang}},\
  }\href {\doibase 10.1063/1.2403848} {\bibfield  {journal} {\bibinfo
  {journal} {The Journal of Chemical Physics}\ }\textbf {\bibinfo {volume}
  {125}},\ \bibinfo {pages} {201102} (\bibinfo {year} {2006})}\BibitemShut
  {NoStop}%
\bibitem [{\citenamefont {Mori-S{\'a}nchez}\ \emph {et~al.}(2008)\citenamefont
  {Mori-S{\'a}nchez}, \citenamefont {Cohen},\ and\ \citenamefont
  {Yang}}]{mori-sanchez_localization_2008}%
  \BibitemOpen
  \bibfield  {author} {\bibinfo {author} {\bibfnamefont {P.}~\bibnamefont
  {Mori-S{\'a}nchez}}, \bibinfo {author} {\bibfnamefont {A.~J.}\ \bibnamefont
  {Cohen}}, \ and\ \bibinfo {author} {\bibfnamefont {W.}~\bibnamefont {Yang}},\
  }\href {\doibase 10.1103/PhysRevLetters100.146401} {\bibfield  {journal}
  {\bibinfo  {journal} {Physical Review Letters}\ }\textbf {\bibinfo {volume}
  {100}},\ \bibinfo {pages} {146401} (\bibinfo {year} {2008})}\BibitemShut
  {NoStop}%
\bibitem [{\citenamefont {Perdew}\ and\ \citenamefont
  {Zunger}(1981)}]{perdew_self-interaction_1981}%
  \BibitemOpen
  \bibfield  {author} {\bibinfo {author} {\bibfnamefont {J.~P.}\ \bibnamefont
  {Perdew}}\ and\ \bibinfo {author} {\bibfnamefont {A.}~\bibnamefont
  {Zunger}},\ }\href {\doibase 10.1103/PhysRevB.23.5048} {\bibfield  {journal}
  {\bibinfo  {journal} {Physical Review B}\ }\textbf {\bibinfo {volume} {23}},\
  \bibinfo {pages} {5048} (\bibinfo {year} {1981})}\BibitemShut {NoStop}%
\bibitem [{\citenamefont {Zheng}\ \emph {et~al.}(2011)\citenamefont {Zheng},
  \citenamefont {Cohen}, \citenamefont {Mori-S\'anchez}, \citenamefont {Hu},\
  and\ \citenamefont {Yang}}]{zheng_improving_2011}%
  \BibitemOpen
  \bibfield  {author} {\bibinfo {author} {\bibfnamefont {X.}~\bibnamefont
  {Zheng}}, \bibinfo {author} {\bibfnamefont {A.~J.}\ \bibnamefont {Cohen}},
  \bibinfo {author} {\bibfnamefont {P.}~\bibnamefont {Mori-S\'anchez}},
  \bibinfo {author} {\bibfnamefont {X.}~\bibnamefont {Hu}}, \ and\ \bibinfo
  {author} {\bibfnamefont {W.}~\bibnamefont {Yang}},\ }\href {\doibase
  10.1103/PhysRevLetters107.026403} {\bibfield  {journal} {\bibinfo  {journal}
  {Physical Review Letters}\ }\textbf {\bibinfo {volume} {107}},\ \bibinfo
  {pages} {026403} (\bibinfo {year} {2011})}\BibitemShut {NoStop}%
\bibitem [{\citenamefont {Kraisler}\ and\ \citenamefont
  {Kronik}(2013)}]{kraisler_piecewise_2013}%
  \BibitemOpen
  \bibfield  {author} {\bibinfo {author} {\bibfnamefont {E.}~\bibnamefont
  {Kraisler}}\ and\ \bibinfo {author} {\bibfnamefont {L.}~\bibnamefont
  {Kronik}},\ }\href {\doibase 10.1103/PhysRevLetters110.126403} {\bibfield
  {journal} {\bibinfo  {journal} {Physical Review Letters}\ }\textbf {\bibinfo
  {volume} {110}},\ \bibinfo {pages} {126403} (\bibinfo {year}
  {2013})}\BibitemShut {NoStop}%
\bibitem [{\citenamefont
  {G{\"o}rling}(2015)}]{gorling_exchange-correlation_2015}%
  \BibitemOpen
  \bibfield  {author} {\bibinfo {author} {\bibfnamefont {A.}~\bibnamefont
  {G{\"o}rling}},\ }\href {\doibase 10.1103/PhysRevB.91.245120} {\bibfield
  {journal} {\bibinfo  {journal} {Physical Review B}\ }\textbf {\bibinfo
  {volume} {91}},\ \bibinfo {pages} {245120} (\bibinfo {year}
  {2015})}\BibitemShut {NoStop}%
\bibitem [{\citenamefont {Li}\ \emph {et~al.}(2015)\citenamefont {Li},
  \citenamefont {Zheng}, \citenamefont {Cohen}, \citenamefont
  {Mori-S\'anchez},\ and\ \citenamefont {Yang}}]{li_local_2015}%
  \BibitemOpen
  \bibfield  {author} {\bibinfo {author} {\bibfnamefont {C.}~\bibnamefont
  {Li}}, \bibinfo {author} {\bibfnamefont {X.}~\bibnamefont {Zheng}}, \bibinfo
  {author} {\bibfnamefont {A.~J.}\ \bibnamefont {Cohen}}, \bibinfo {author}
  {\bibfnamefont {P.}~\bibnamefont {Mori-S\'anchez}}, \ and\ \bibinfo {author}
  {\bibfnamefont {W.}~\bibnamefont {Yang}},\ }\href {\doibase
  10.1103/PhysRevLetters114.053001} {\bibfield  {journal} {\bibinfo  {journal}
  {Physical Review Letters}\ }\textbf {\bibinfo {volume} {114}},\ \bibinfo
  {pages} {053001} (\bibinfo {year} {2015})}\BibitemShut {NoStop}%
\bibitem [{\citenamefont {Stein}\ \emph {et~al.}(2010)\citenamefont {Stein},
  \citenamefont {Eisenberg}, \citenamefont {Kronik},\ and\ \citenamefont
  {Baer}}]{stein_fundamental_2010}%
  \BibitemOpen
  \bibfield  {author} {\bibinfo {author} {\bibfnamefont {T.}~\bibnamefont
  {Stein}}, \bibinfo {author} {\bibfnamefont {H.}~\bibnamefont {Eisenberg}},
  \bibinfo {author} {\bibfnamefont {L.}~\bibnamefont {Kronik}}, \ and\ \bibinfo
  {author} {\bibfnamefont {R.}~\bibnamefont {Baer}},\ }\href {\doibase
  10.1103/PhysRevLetters105.266802} {\bibfield  {journal} {\bibinfo  {journal}
  {Physical Review Letters}\ }\textbf {\bibinfo {volume} {105}},\ \bibinfo
  {pages} {266802} (\bibinfo {year} {2010})}\BibitemShut {NoStop}%
\bibitem [{\citenamefont {Kronik}\ \emph {et~al.}(2012)\citenamefont {Kronik},
  \citenamefont {Stein}, \citenamefont {Refaely-Abramson},\ and\ \citenamefont
  {Baer}}]{kronik_excitation_2012}%
  \BibitemOpen
  \bibfield  {author} {\bibinfo {author} {\bibfnamefont {L.}~\bibnamefont
  {Kronik}}, \bibinfo {author} {\bibfnamefont {T.}~\bibnamefont {Stein}},
  \bibinfo {author} {\bibfnamefont {S.}~\bibnamefont {Refaely-Abramson}}, \
  and\ \bibinfo {author} {\bibfnamefont {R.}~\bibnamefont {Baer}},\ }\href
  {\doibase 10.1021/ct2009363} {\bibfield  {journal} {\bibinfo  {journal}
  {Journal of Chemical Theory and Computation}\ }\textbf {\bibinfo {volume}
  {8}},\ \bibinfo {pages} {1515} (\bibinfo {year} {2012})}\BibitemShut
  {NoStop}%
\bibitem [{\citenamefont {Refaely-Abramson}\ \emph {et~al.}(2012)\citenamefont
  {Refaely-Abramson}, \citenamefont {Sharifzadeh}, \citenamefont {Govind},
  \citenamefont {Autschbach}, \citenamefont {Neaton}, \citenamefont {Baer},\
  and\ \citenamefont {Kronik}}]{refaely-abramson_quasiparticle_2012}%
  \BibitemOpen
  \bibfield  {author} {\bibinfo {author} {\bibfnamefont {S.}~\bibnamefont
  {Refaely-Abramson}}, \bibinfo {author} {\bibfnamefont {S.}~\bibnamefont
  {Sharifzadeh}}, \bibinfo {author} {\bibfnamefont {N.}~\bibnamefont {Govind}},
  \bibinfo {author} {\bibfnamefont {J.}~\bibnamefont {Autschbach}}, \bibinfo
  {author} {\bibfnamefont {J.~B.}\ \bibnamefont {Neaton}}, \bibinfo {author}
  {\bibfnamefont {R.}~\bibnamefont {Baer}}, \ and\ \bibinfo {author}
  {\bibfnamefont {L.}~\bibnamefont {Kronik}},\ }\href@noop {} {\bibfield
  {journal} {\bibinfo  {journal} {Physical Review Lettersrs}\ }\textbf
  {\bibinfo {volume} {109}},\ \bibinfo {pages} {226405} (\bibinfo {year}
  {2012})}\BibitemShut {NoStop}%
\bibitem [{\citenamefont {Shimazaki}\ and\ \citenamefont
  {Asai}(2008)}]{shimazaki_band_2008}%
  \BibitemOpen
  \bibfield  {author} {\bibinfo {author} {\bibfnamefont {T.}~\bibnamefont
  {Shimazaki}}\ and\ \bibinfo {author} {\bibfnamefont {Y.}~\bibnamefont
  {Asai}},\ }\href {\doibase https://doi.org/10.1016/j.cplett.2008.10.012}
  {\bibfield  {journal} {\bibinfo  {journal} {Chemical Physics Lettersrs}\
  }\textbf {\bibinfo {volume} {466}},\ \bibinfo {pages} {91 } (\bibinfo {year}
  {2008})}\BibitemShut {NoStop}%
\bibitem [{\citenamefont {Skone}\ \emph {et~al.}(2014)\citenamefont {Skone},
  \citenamefont {Govoni},\ and\ \citenamefont
  {Galli}}]{skone_self-consistent_2014}%
  \BibitemOpen
  \bibfield  {author} {\bibinfo {author} {\bibfnamefont {J.~H.}\ \bibnamefont
  {Skone}}, \bibinfo {author} {\bibfnamefont {M.}~\bibnamefont {Govoni}}, \
  and\ \bibinfo {author} {\bibfnamefont {G.}~\bibnamefont {Galli}},\ }\href
  {\doibase 10.1103/PhysRevB.89.195112} {\bibfield  {journal} {\bibinfo
  {journal} {Physical Review B}\ }\textbf {\bibinfo {volume} {89}},\ \bibinfo
  {pages} {195112} (\bibinfo {year} {2014})}\BibitemShut {NoStop}%
\bibitem [{\citenamefont {Brawand}\ \emph {et~al.}(2016)\citenamefont
  {Brawand}, \citenamefont {V{\"o}r{\"o}s}, \citenamefont {Govoni},\ and\
  \citenamefont {Galli}}]{brawand_generalization_2016}%
  \BibitemOpen
  \bibfield  {author} {\bibinfo {author} {\bibfnamefont {N.~P.}\ \bibnamefont
  {Brawand}}, \bibinfo {author} {\bibfnamefont {M.}~\bibnamefont
  {V{\"o}r{\"o}s}}, \bibinfo {author} {\bibfnamefont {M.}~\bibnamefont
  {Govoni}}, \ and\ \bibinfo {author} {\bibfnamefont {G.}~\bibnamefont
  {Galli}},\ }\href {\doibase 10.1103/PhysRevX.6.041002} {\bibfield  {journal}
  {\bibinfo  {journal} {Physical Review X}\ }\textbf {\bibinfo {volume} {6}},\
  \bibinfo {pages} {041002} (\bibinfo {year} {2016})}\BibitemShut {NoStop}%
\bibitem [{\citenamefont {Dabo}\ \emph {et~al.}(2009)\citenamefont {Dabo},
  \citenamefont {Cococcioni},\ and\ \citenamefont
  {Marzari}}]{dabo_non-koopmans_2009}%
  \BibitemOpen
  \bibfield  {author} {\bibinfo {author} {\bibfnamefont {I.}~\bibnamefont
  {Dabo}}, \bibinfo {author} {\bibfnamefont {M.}~\bibnamefont {Cococcioni}}, \
  and\ \bibinfo {author} {\bibfnamefont {N.}~\bibnamefont {Marzari}},\ }\href
  {http://arxiv.org/abs/0901.2637} {\bibfield  {journal} {\bibinfo  {journal}
  {arXiv:0901.2637 [cond-mat]}\ } (\bibinfo {year} {2009})},\ \bibinfo {note}
  {arXiv: 0901.2637}\BibitemShut {NoStop}%
\bibitem [{\citenamefont {Dabo}\ \emph {et~al.}(2010)\citenamefont {Dabo},
  \citenamefont {Ferretti}, \citenamefont {Poilvert}, \citenamefont {Li},
  \citenamefont {Marzari},\ and\ \citenamefont
  {Cococcioni}}]{dabo_koopmans_2010}%
  \BibitemOpen
  \bibfield  {author} {\bibinfo {author} {\bibfnamefont {I.}~\bibnamefont
  {Dabo}}, \bibinfo {author} {\bibfnamefont {A.}~\bibnamefont {Ferretti}},
  \bibinfo {author} {\bibfnamefont {N.}~\bibnamefont {Poilvert}}, \bibinfo
  {author} {\bibfnamefont {Y.}~\bibnamefont {Li}}, \bibinfo {author}
  {\bibfnamefont {N.}~\bibnamefont {Marzari}}, \ and\ \bibinfo {author}
  {\bibfnamefont {M.}~\bibnamefont {Cococcioni}},\ }\href {\doibase
  10.1103/PhysRevB.82.115121} {\bibfield  {journal} {\bibinfo  {journal}
  {Physical Review B}\ }\textbf {\bibinfo {volume} {82}},\ \bibinfo {pages}
  {115121} (\bibinfo {year} {2010})}\BibitemShut {NoStop}%
\bibitem [{\citenamefont {Dabo}\ \emph {et~al.}(2013)\citenamefont {Dabo},
  \citenamefont {Ferretti}, \citenamefont {Borghi}, \citenamefont {Nguyen},
  \citenamefont {Park}, \citenamefont {Cococcioni},\ and\ \citenamefont
  {Marzari}}]{dabo_piecewise_2013}%
  \BibitemOpen
  \bibfield  {author} {\bibinfo {author} {\bibfnamefont {I.}~\bibnamefont
  {Dabo}}, \bibinfo {author} {\bibfnamefont {A.}~\bibnamefont {Ferretti}},
  \bibinfo {author} {\bibfnamefont {G.}~\bibnamefont {Borghi}}, \bibinfo
  {author} {\bibfnamefont {N.~L.}\ \bibnamefont {Nguyen}}, \bibinfo {author}
  {\bibfnamefont {C.-H.}\ \bibnamefont {Park}}, \bibinfo {author}
  {\bibfnamefont {M.}~\bibnamefont {Cococcioni}}, \ and\ \bibinfo {author}
  {\bibfnamefont {N.}~\bibnamefont {Marzari}},\ }\href
  {http://psi-k.net/download/highlights/Highlight_119.pdf} {\bibfield
  {journal} {\bibinfo  {journal} {Psi-K Newsletter}\ }\textbf {\bibinfo
  {volume} {119}} (\bibinfo {year} {2013})}\BibitemShut {NoStop}%
\bibitem [{\citenamefont {Dabo}\ \emph {et~al.}(2012)\citenamefont {Dabo},
  \citenamefont {Ferretti}, \citenamefont {Park}, \citenamefont {Poilvert},
  \citenamefont {Li}, \citenamefont {Cococcioni},\ and\ \citenamefont
  {Marzari}}]{dabo_donor_2012}%
  \BibitemOpen
  \bibfield  {author} {\bibinfo {author} {\bibfnamefont {I.}~\bibnamefont
  {Dabo}}, \bibinfo {author} {\bibfnamefont {A.}~\bibnamefont {Ferretti}},
  \bibinfo {author} {\bibfnamefont {C.-H.}\ \bibnamefont {Park}}, \bibinfo
  {author} {\bibfnamefont {N.}~\bibnamefont {Poilvert}}, \bibinfo {author}
  {\bibfnamefont {Y.}~\bibnamefont {Li}}, \bibinfo {author} {\bibfnamefont
  {M.}~\bibnamefont {Cococcioni}}, \ and\ \bibinfo {author} {\bibfnamefont
  {N.}~\bibnamefont {Marzari}},\ }\href {\doibase 10.1039/C2CP43491A}
  {\bibfield  {journal} {\bibinfo  {journal} {Phys. Chem. Chem. Phys.}\
  }\textbf {\bibinfo {volume} {15}},\ \bibinfo {pages} {685} (\bibinfo {year}
  {2012})}\BibitemShut {NoStop}%
\bibitem [{\citenamefont {Borghi}\ \emph {et~al.}(2014)\citenamefont {Borghi},
  \citenamefont {Ferretti}, \citenamefont {Nguyen}, \citenamefont {Dabo},\ and\
  \citenamefont {Marzari}}]{borghi_koopmans-compliant_2014}%
  \BibitemOpen
  \bibfield  {author} {\bibinfo {author} {\bibfnamefont {G.}~\bibnamefont
  {Borghi}}, \bibinfo {author} {\bibfnamefont {A.}~\bibnamefont {Ferretti}},
  \bibinfo {author} {\bibfnamefont {N.~L.}\ \bibnamefont {Nguyen}}, \bibinfo
  {author} {\bibfnamefont {I.}~\bibnamefont {Dabo}}, \ and\ \bibinfo {author}
  {\bibfnamefont {N.}~\bibnamefont {Marzari}},\ }\href {\doibase
  10.1103/PhysRevB.90.075135} {\bibfield  {journal} {\bibinfo  {journal}
  {Physical Review B}\ }\textbf {\bibinfo {volume} {90}},\ \bibinfo {pages}
  {075135} (\bibinfo {year} {2014})}\BibitemShut {NoStop}%
\bibitem [{\citenamefont {Ferretti}\ \emph {et~al.}(2014)\citenamefont
  {Ferretti}, \citenamefont {Dabo}, \citenamefont {Cococcioni},\ and\
  \citenamefont {Marzari}}]{ferretti_bridging_2014}%
  \BibitemOpen
  \bibfield  {author} {\bibinfo {author} {\bibfnamefont {A.}~\bibnamefont
  {Ferretti}}, \bibinfo {author} {\bibfnamefont {I.}~\bibnamefont {Dabo}},
  \bibinfo {author} {\bibfnamefont {M.}~\bibnamefont {Cococcioni}}, \ and\
  \bibinfo {author} {\bibfnamefont {N.}~\bibnamefont {Marzari}},\ }\href
  {\doibase 10.1103/PhysRevB.89.195134} {\bibfield  {journal} {\bibinfo
  {journal} {Physical Review B}\ }\textbf {\bibinfo {volume} {89}},\ \bibinfo
  {pages} {195134} (\bibinfo {year} {2014})}\BibitemShut {NoStop}%
\bibitem [{\citenamefont {Gatti}\ \emph {et~al.}(2007)\citenamefont {Gatti},
  \citenamefont {Olevano}, \citenamefont {Reining},\ and\ \citenamefont
  {Tokatly}}]{gatti_transforming_2007}%
  \BibitemOpen
  \bibfield  {author} {\bibinfo {author} {\bibfnamefont {M.}~\bibnamefont
  {Gatti}}, \bibinfo {author} {\bibfnamefont {V.}~\bibnamefont {Olevano}},
  \bibinfo {author} {\bibfnamefont {L.}~\bibnamefont {Reining}}, \ and\
  \bibinfo {author} {\bibfnamefont {I.~V.}\ \bibnamefont {Tokatly}},\ }\href
  {\doibase 10.1103/PhysRevLetters99.057401} {\bibfield  {journal} {\bibinfo
  {journal} {Physical Review Letters}\ }\textbf {\bibinfo {volume} {99}},\
  \bibinfo {pages} {057401} (\bibinfo {year} {2007})}\BibitemShut {NoStop}%
\bibitem [{\citenamefont {Heaton}\ \emph {et~al.}(1983)\citenamefont {Heaton},
  \citenamefont {Harrison},\ and\ \citenamefont
  {Lin}}]{heaton_self-interaction_1983}%
  \BibitemOpen
  \bibfield  {author} {\bibinfo {author} {\bibfnamefont {R.~A.}\ \bibnamefont
  {Heaton}}, \bibinfo {author} {\bibfnamefont {J.~G.}\ \bibnamefont
  {Harrison}}, \ and\ \bibinfo {author} {\bibfnamefont {C.~C.}\ \bibnamefont
  {Lin}},\ }\href {\doibase 10.1103/PhysRevB.28.5992} {\bibfield  {journal}
  {\bibinfo  {journal} {Physical Review B}\ }\textbf {\bibinfo {volume} {28}},\
  \bibinfo {pages} {5992} (\bibinfo {year} {1983})}\BibitemShut {NoStop}%
\bibitem [{\citenamefont {Pederson}\ \emph {et~al.}(1984)\citenamefont
  {Pederson}, \citenamefont {Heaton},\ and\ \citenamefont
  {Lin}}]{pederson_localdensity_1984}%
  \BibitemOpen
  \bibfield  {author} {\bibinfo {author} {\bibfnamefont {M.~R.}\ \bibnamefont
  {Pederson}}, \bibinfo {author} {\bibfnamefont {R.~A.}\ \bibnamefont
  {Heaton}}, \ and\ \bibinfo {author} {\bibfnamefont {C.~C.}\ \bibnamefont
  {Lin}},\ }\href {\doibase 10.1063/1.446959} {\bibfield  {journal} {\bibinfo
  {journal} {The Journal of Chemical Physics}\ }\textbf {\bibinfo {volume}
  {80}},\ \bibinfo {pages} {1972} (\bibinfo {year} {1984})}\BibitemShut
  {NoStop}%
\bibitem [{\citenamefont {Pederson}\ \emph {et~al.}(1985)\citenamefont
  {Pederson}, \citenamefont {Heaton},\ and\ \citenamefont
  {Lin}}]{pederson_densityfunctional_1985}%
  \BibitemOpen
  \bibfield  {author} {\bibinfo {author} {\bibfnamefont {M.~R.}\ \bibnamefont
  {Pederson}}, \bibinfo {author} {\bibfnamefont {R.~A.}\ \bibnamefont
  {Heaton}}, \ and\ \bibinfo {author} {\bibfnamefont {C.~C.}\ \bibnamefont
  {Lin}},\ }\href {\doibase 10.1063/1.448266} {\bibfield  {journal} {\bibinfo
  {journal} {The Journal of Chemical Physics}\ }\textbf {\bibinfo {volume}
  {82}},\ \bibinfo {pages} {2688} (\bibinfo {year} {1985})}\BibitemShut
  {NoStop}%
\bibitem [{\citenamefont {Pederson}\ and\ \citenamefont
  {Lin}(1988)}]{pederson_localized_1988}%
  \BibitemOpen
  \bibfield  {author} {\bibinfo {author} {\bibfnamefont {M.~R.}\ \bibnamefont
  {Pederson}}\ and\ \bibinfo {author} {\bibfnamefont {C.~C.}\ \bibnamefont
  {Lin}},\ }\href {\doibase 10.1063/1.454104} {\bibfield  {journal} {\bibinfo
  {journal} {The Journal of Chemical Physics}\ }\textbf {\bibinfo {volume}
  {88}},\ \bibinfo {pages} {1807} (\bibinfo {year} {1988})}\BibitemShut
  {NoStop}%
\bibitem [{\citenamefont {Lehtola}\ and\ \citenamefont
  {J{\'o}nsson}(2014)}]{lehtola_variational_2014}%
  \BibitemOpen
  \bibfield  {author} {\bibinfo {author} {\bibfnamefont {S.}~\bibnamefont
  {Lehtola}}\ and\ \bibinfo {author} {\bibfnamefont {H.}~\bibnamefont
  {J{\'o}nsson}},\ }\href {\doibase 10.1021/ct500637x} {\bibfield  {journal}
  {\bibinfo  {journal} {J. Chem. Theory Comput.}\ }\textbf {\bibinfo {volume}
  {10}},\ \bibinfo {pages} {5324} (\bibinfo {year} {2014})}\BibitemShut
  {NoStop}%
\bibitem [{\citenamefont {Boys}(1960)}]{boys_construction_1960}%
  \BibitemOpen
  \bibfield  {author} {\bibinfo {author} {\bibfnamefont {S.~F.}\ \bibnamefont
  {Boys}},\ }\href {\doibase 10.1103/RevModPhys.32.296} {\bibfield  {journal}
  {\bibinfo  {journal} {Reviews of Modern Physics}\ }\textbf {\bibinfo {volume}
  {32}},\ \bibinfo {pages} {296} (\bibinfo {year} {1960})}\BibitemShut
  {NoStop}%
\bibitem [{\citenamefont {Marzari}\ \emph {et~al.}(2012)\citenamefont
  {Marzari}, \citenamefont {Mostofi}, \citenamefont {Yates}, \citenamefont
  {Souza},\ and\ \citenamefont {Vanderbilt}}]{marzari_maximally_2012}%
  \BibitemOpen
  \bibfield  {author} {\bibinfo {author} {\bibfnamefont {N.}~\bibnamefont
  {Marzari}}, \bibinfo {author} {\bibfnamefont {A.~A.}\ \bibnamefont
  {Mostofi}}, \bibinfo {author} {\bibfnamefont {J.~R.}\ \bibnamefont {Yates}},
  \bibinfo {author} {\bibfnamefont {I.}~\bibnamefont {Souza}}, \ and\ \bibinfo
  {author} {\bibfnamefont {D.}~\bibnamefont {Vanderbilt}},\ }\href {\doibase
  10.1103/RevModPhys.84.1419} {\bibfield  {journal} {\bibinfo  {journal}
  {Reviews of Modern Physics}\ }\textbf {\bibinfo {volume} {84}},\ \bibinfo
  {pages} {1419} (\bibinfo {year} {2012})}\BibitemShut {NoStop}%
\bibitem [{\citenamefont {Goedecker}\ and\ \citenamefont
  {Umrigar}(1997)}]{goedecker_critical_1997}%
  \BibitemOpen
  \bibfield  {author} {\bibinfo {author} {\bibfnamefont {S.}~\bibnamefont
  {Goedecker}}\ and\ \bibinfo {author} {\bibfnamefont {C.~J.}\ \bibnamefont
  {Umrigar}},\ }\href {\doibase 10.1103/PhysRevA.55.1765} {\bibfield  {journal}
  {\bibinfo  {journal} {Physical Review A}\ }\textbf {\bibinfo {volume} {55}},\
  \bibinfo {pages} {1765} (\bibinfo {year} {1997})}\BibitemShut {NoStop}%
\bibitem [{\citenamefont {Stengel}\ and\ \citenamefont
  {Spaldin}(2008)}]{stengel_self-interaction_2008}%
  \BibitemOpen
  \bibfield  {author} {\bibinfo {author} {\bibfnamefont {M.}~\bibnamefont
  {Stengel}}\ and\ \bibinfo {author} {\bibfnamefont {N.~A.}\ \bibnamefont
  {Spaldin}},\ }\href {\doibase 10.1103/PhysRevB.77.155106} {\bibfield
  {journal} {\bibinfo  {journal} {Physical Review B}\ }\textbf {\bibinfo
  {volume} {77}},\ \bibinfo {pages} {155106} (\bibinfo {year}
  {2008})}\BibitemShut {NoStop}%
\bibitem [{\citenamefont {Borghi}\ \emph {et~al.}(2015)\citenamefont {Borghi},
  \citenamefont {Park}, \citenamefont {Nguyen}, \citenamefont {Ferretti},\ and\
  \citenamefont {Marzari}}]{borghi_variational_2015}%
  \BibitemOpen
  \bibfield  {author} {\bibinfo {author} {\bibfnamefont {G.}~\bibnamefont
  {Borghi}}, \bibinfo {author} {\bibfnamefont {C.-H.}\ \bibnamefont {Park}},
  \bibinfo {author} {\bibfnamefont {N.~L.}\ \bibnamefont {Nguyen}}, \bibinfo
  {author} {\bibfnamefont {A.}~\bibnamefont {Ferretti}}, \ and\ \bibinfo
  {author} {\bibfnamefont {N.}~\bibnamefont {Marzari}},\ }\href {\doibase
  10.1103/PhysRevB.91.155112} {\bibfield  {journal} {\bibinfo  {journal}
  {Physical Review B}\ }\textbf {\bibinfo {volume} {91}},\ \bibinfo {pages}
  {155112} (\bibinfo {year} {2015})}\BibitemShut {NoStop}%
\bibitem [{\citenamefont {K{\"o}rzd{\"o}rfer}\ \emph
  {et~al.}(2008)\citenamefont {K{\"o}rzd{\"o}rfer}, \citenamefont
  {K{\"u}mmel},\ and\ \citenamefont
  {Mundt}}]{korzdorfer_self-interaction_2008}%
  \BibitemOpen
  \bibfield  {author} {\bibinfo {author} {\bibfnamefont {T.}~\bibnamefont
  {K{\"o}rzd{\"o}rfer}}, \bibinfo {author} {\bibfnamefont {S.}~\bibnamefont
  {K{\"u}mmel}}, \ and\ \bibinfo {author} {\bibfnamefont {M.}~\bibnamefont
  {Mundt}},\ }\href {\doibase 10.1063/1.2944272} {\bibfield  {journal}
  {\bibinfo  {journal} {The Journal of Chemical Physics}\ }\textbf {\bibinfo
  {volume} {129}},\ \bibinfo {pages} {014110} (\bibinfo {year}
  {2008})}\BibitemShut {NoStop}%
\bibitem [{\citenamefont {Vydrov}\ \emph {et~al.}(2007)\citenamefont {Vydrov},
  \citenamefont {Scuseria},\ and\ \citenamefont {Perdew}}]{vydrov_tests_2007}%
  \BibitemOpen
  \bibfield  {author} {\bibinfo {author} {\bibfnamefont {O.~A.}\ \bibnamefont
  {Vydrov}}, \bibinfo {author} {\bibfnamefont {G.~E.}\ \bibnamefont
  {Scuseria}}, \ and\ \bibinfo {author} {\bibfnamefont {J.~P.}\ \bibnamefont
  {Perdew}},\ }\href {\doibase 10.1063/1.2723119} {\bibfield  {journal}
  {\bibinfo  {journal} {The Journal of Chemical Physics}\ }\textbf {\bibinfo
  {volume} {126}},\ \bibinfo {pages} {154109} (\bibinfo {year}
  {2007})}\BibitemShut {NoStop}%
\bibitem [{\citenamefont {Vydrov}\ and\ \citenamefont
  {Scuseria}(2005)}]{vydrov_ionization_2005}%
  \BibitemOpen
  \bibfield  {author} {\bibinfo {author} {\bibfnamefont {O.~A.}\ \bibnamefont
  {Vydrov}}\ and\ \bibinfo {author} {\bibfnamefont {G.~E.}\ \bibnamefont
  {Scuseria}},\ }\href {\doibase 10.1063/1.1897378} {\bibfield  {journal}
  {\bibinfo  {journal} {The Journal of Chemical Physics}\ }\textbf {\bibinfo
  {volume} {122}},\ \bibinfo {pages} {184107} (\bibinfo {year}
  {2005})}\BibitemShut {NoStop}%
\bibitem [{\citenamefont {Nguyen}\ \emph {et~al.}(2015)\citenamefont {Nguyen},
  \citenamefont {Borghi}, \citenamefont {Ferretti}, \citenamefont {Dabo},\ and\
  \citenamefont {Marzari}}]{nguyen_first-principles_2015}%
  \BibitemOpen
  \bibfield  {author} {\bibinfo {author} {\bibfnamefont {N.~L.}\ \bibnamefont
  {Nguyen}}, \bibinfo {author} {\bibfnamefont {G.}~\bibnamefont {Borghi}},
  \bibinfo {author} {\bibfnamefont {A.}~\bibnamefont {Ferretti}}, \bibinfo
  {author} {\bibfnamefont {I.}~\bibnamefont {Dabo}}, \ and\ \bibinfo {author}
  {\bibfnamefont {N.}~\bibnamefont {Marzari}},\ }\href {\doibase
  10.1103/PhysRevLetters114.166405} {\bibfield  {journal} {\bibinfo  {journal}
  {Physical Review Letters}\ }\textbf {\bibinfo {volume} {114}},\ \bibinfo
  {pages} {166405} (\bibinfo {year} {2015})}\BibitemShut {NoStop}%
\bibitem [{\citenamefont {Nguyen}\ \emph {et~al.}(2016)\citenamefont {Nguyen},
  \citenamefont {Borghi}, \citenamefont {Ferretti},\ and\ \citenamefont
  {Marzari}}]{nguyen_first-principles_2016}%
  \BibitemOpen
  \bibfield  {author} {\bibinfo {author} {\bibfnamefont {N.~L.}\ \bibnamefont
  {Nguyen}}, \bibinfo {author} {\bibfnamefont {G.}~\bibnamefont {Borghi}},
  \bibinfo {author} {\bibfnamefont {A.}~\bibnamefont {Ferretti}}, \ and\
  \bibinfo {author} {\bibfnamefont {N.}~\bibnamefont {Marzari}},\ }\href
  {\doibase 10.1021/acs.jctc.6b00145} {\bibfield  {journal} {\bibinfo
  {journal} {J. Chem. Theory Comput.}\ }\textbf {\bibinfo {volume} {12}},\
  \bibinfo {pages} {3948} (\bibinfo {year} {2016})}\BibitemShut {NoStop}%
\bibitem [{\citenamefont {Nguyen}\ \emph {et~al.}(2017)\citenamefont {Nguyen},
  \citenamefont {Colonna}, \citenamefont {Ferretti},\ and\ \citenamefont
  {Marzari}}]{nguyen_koopmans-compliant_2017}%
  \BibitemOpen
  \bibfield  {author} {\bibinfo {author} {\bibfnamefont {N.~L.}\ \bibnamefont
  {Nguyen}}, \bibinfo {author} {\bibfnamefont {N.}~\bibnamefont {Colonna}},
  \bibinfo {author} {\bibfnamefont {A.}~\bibnamefont {Ferretti}}, \ and\
  \bibinfo {author} {\bibfnamefont {N.}~\bibnamefont {Marzari}},\ }\href
  {http://arxiv.org/abs/1708.08518} {\bibfield  {journal} {\bibinfo  {journal}
  {arXiv:1708.08518 [cond-mat]}\ } (\bibinfo {year} {2017})},\ \bibinfo {note}
  {arXiv: 1708.08518}\BibitemShut {NoStop}%
\bibitem [{\citenamefont {Ma}\ and\ \citenamefont
  {Wang}(2016)}]{ma_using_2016}%
  \BibitemOpen
  \bibfield  {author} {\bibinfo {author} {\bibfnamefont {J.}~\bibnamefont
  {Ma}}\ and\ \bibinfo {author} {\bibfnamefont {L.-W.}\ \bibnamefont {Wang}},\
  }\href {\doibase 10.1038/srep24924} {\bibfield  {journal} {\bibinfo
  {journal} {Scientific Reports}\ }\textbf {\bibinfo {volume} {6}},\ \bibinfo
  {pages} {24924} (\bibinfo {year} {2016})}\BibitemShut {NoStop}%
\bibitem [{\citenamefont {Martin}\ \emph {et~al.}(2016)\citenamefont {Martin},
  \citenamefont {Reining},\ and\ \citenamefont
  {Ceperley}}]{martin_interacting_2016}%
  \BibitemOpen
  \bibfield  {author} {\bibinfo {author} {\bibfnamefont {R.~M.}\ \bibnamefont
  {Martin}}, \bibinfo {author} {\bibfnamefont {L.}~\bibnamefont {Reining}}, \
  and\ \bibinfo {author} {\bibfnamefont {D.~M.}\ \bibnamefont {Ceperley}},\
  }\href@noop {} {\emph {\bibinfo {title} {Interacting {Electrons}}}}\
  (\bibinfo  {publisher} {Cambridge University Press},\ \bibinfo {year}
  {2016})\BibitemShut {NoStop}%
\bibitem [{\citenamefont {Gross}\ and\ \citenamefont
  {Kohn}(1985)}]{gross_local_1985}%
  \BibitemOpen
  \bibfield  {author} {\bibinfo {author} {\bibfnamefont {E.~K.~U.}\
  \bibnamefont {Gross}}\ and\ \bibinfo {author} {\bibfnamefont
  {W.}~\bibnamefont {Kohn}},\ }\href {\doibase 10.1103/PhysRevLetters55.2850}
  {\bibfield  {journal} {\bibinfo  {journal} {Physical Review Letters}\
  }\textbf {\bibinfo {volume} {55}},\ \bibinfo {pages} {2850} (\bibinfo {year}
  {1985})}\BibitemShut {NoStop}%
\bibitem [{\citenamefont {Dederichs}\ \emph {et~al.}(1984)\citenamefont
  {Dederichs}, \citenamefont {Bl{\"u}gel}, \citenamefont {Zeller},\ and\
  \citenamefont {Akai}}]{dederichs_ground_1984}%
  \BibitemOpen
  \bibfield  {author} {\bibinfo {author} {\bibfnamefont {P.~H.}\ \bibnamefont
  {Dederichs}}, \bibinfo {author} {\bibfnamefont {S.}~\bibnamefont
  {Bl{\"u}gel}}, \bibinfo {author} {\bibfnamefont {R.}~\bibnamefont {Zeller}},
  \ and\ \bibinfo {author} {\bibfnamefont {H.}~\bibnamefont {Akai}},\ }\href
  {\doibase 10.1103/PhysRevLetters53.2512} {\bibfield  {journal} {\bibinfo
  {journal} {Physical Review Letters}\ }\textbf {\bibinfo {volume} {53}},\
  \bibinfo {pages} {2512} (\bibinfo {year} {1984})}\BibitemShut {NoStop}%
\bibitem [{\citenamefont {Anisimov}\ and\ \citenamefont
  {Gunnarsson}(1991)}]{anisimov_density-functional_1991}%
  \BibitemOpen
  \bibfield  {author} {\bibinfo {author} {\bibfnamefont {V.~I.}\ \bibnamefont
  {Anisimov}}\ and\ \bibinfo {author} {\bibfnamefont {O.}~\bibnamefont
  {Gunnarsson}},\ }\href {\doibase 10.1103/PhysRevB.43.7570} {\bibfield
  {journal} {\bibinfo  {journal} {Physical Review B}\ }\textbf {\bibinfo
  {volume} {43}},\ \bibinfo {pages} {7570} (\bibinfo {year}
  {1991})}\BibitemShut {NoStop}%
\bibitem [{\citenamefont {Hybertsen}\ \emph {et~al.}(1989)\citenamefont
  {Hybertsen}, \citenamefont {Schl{\"u}ter},\ and\ \citenamefont
  {Christensen}}]{hybertsen_calculation_1989}%
  \BibitemOpen
  \bibfield  {author} {\bibinfo {author} {\bibfnamefont {M.~S.}\ \bibnamefont
  {Hybertsen}}, \bibinfo {author} {\bibfnamefont {M.}~\bibnamefont
  {Schl{\"u}ter}}, \ and\ \bibinfo {author} {\bibfnamefont {N.~E.}\
  \bibnamefont {Christensen}},\ }\href {\doibase 10.1103/PhysRevB.39.9028}
  {\bibfield  {journal} {\bibinfo  {journal} {Physical Review B}\ }\textbf
  {\bibinfo {volume} {39}},\ \bibinfo {pages} {9028} (\bibinfo {year}
  {1989})}\BibitemShut {NoStop}%
\bibitem [{\citenamefont {McMahan}\ \emph {et~al.}(1988)\citenamefont
  {McMahan}, \citenamefont {Martin},\ and\ \citenamefont
  {Satpathy}}]{mcmahan_calculated_1988}%
  \BibitemOpen
  \bibfield  {author} {\bibinfo {author} {\bibfnamefont {A.~K.}\ \bibnamefont
  {McMahan}}, \bibinfo {author} {\bibfnamefont {R.~M.}\ \bibnamefont {Martin}},
  \ and\ \bibinfo {author} {\bibfnamefont {S.}~\bibnamefont {Satpathy}},\
  }\href {\doibase 10.1103/PhysRevB.38.6650} {\bibfield  {journal} {\bibinfo
  {journal} {Physical Review B}\ }\textbf {\bibinfo {volume} {38}},\ \bibinfo
  {pages} {6650} (\bibinfo {year} {1988})}\BibitemShut {NoStop}%
\bibitem [{\citenamefont {Janak}(1978)}]{janak_proof_1978}%
  \BibitemOpen
  \bibfield  {author} {\bibinfo {author} {\bibfnamefont {J.~F.}\ \bibnamefont
  {Janak}},\ }\href {\doibase 10.1103/PhysRevB.18.7165} {\bibfield  {journal}
  {\bibinfo  {journal} {Physical Review B}\ }\textbf {\bibinfo {volume} {18}},\
  \bibinfo {pages} {7165} (\bibinfo {year} {1978})}\BibitemShut {NoStop}%
\bibitem [{Note1()}]{Note1}%
  \BibitemOpen
  \bibinfo {note} {This restriction can be however released (see Supporting
  Information) and the final result of the derivation [Eq.~(\ref {eq.2urPiKi})]
  apply to any equivalent representation of the KS orbitals, i.e. to any set of
  orbitals related to the KS one by a unitary transformation.}\BibitemShut
  {Stop}%
\bibitem [{\citenamefont {Hellmann}(1944)}]{hellmann_einfuhrung_1944}%
  \BibitemOpen
  \bibfield  {author} {\bibinfo {author} {\bibfnamefont {H.}~\bibnamefont
  {Hellmann}},\ } {\bibinfo {title}
  {Einf{\"u}hrung in die {Quantenchemie}}}\ (\bibinfo  {publisher} {J.W.
  Edwards},\ \bibinfo {address} {Ann Arbor, Mich},\ \bibinfo {year} {1944})\
  \bibinfo {note} {open Library ID: OL21481721M}\BibitemShut {NoStop}%
\bibitem [{\citenamefont {Feynman}(1939)}]{feynman_forces_1939}%
  \BibitemOpen
  \bibfield  {author} {\bibinfo {author} {\bibfnamefont {R.~P.}\ \bibnamefont
  {Feynman}},\ }\href {\doibase 10.1103/PhysRev.56.340} {\bibfield  {journal}
  {\bibinfo  {journal} {Physical Review}\ }\textbf {\bibinfo {volume} {56}},\
  \bibinfo {pages} {340} (\bibinfo {year} {1939})}\BibitemShut {NoStop}%
\bibitem [{\citenamefont {Petersilka}\ \emph {et~al.}(1996)\citenamefont
  {Petersilka}, \citenamefont {Gossmann},\ and\ \citenamefont
  {Gross}}]{petersilka_excitation_1996}%
  \BibitemOpen
  \bibfield  {author} {\bibinfo {author} {\bibfnamefont {M.}~\bibnamefont
  {Petersilka}}, \bibinfo {author} {\bibfnamefont {U.~J.}\ \bibnamefont
  {Gossmann}}, \ and\ \bibinfo {author} {\bibfnamefont {E.~K.~U.}\ \bibnamefont
  {Gross}},\ }\href {\doibase 10.1103/PhysRevLetters76.1212} {\bibfield
  {journal} {\bibinfo  {journal} {Physical Review Letters}\ }\textbf {\bibinfo
  {volume} {76}},\ \bibinfo {pages} {1212} (\bibinfo {year}
  {1996})}\BibitemShut {NoStop}%
\bibitem [{\citenamefont {Nalewajski}(1996)}]{nalewajski_density_1996}%
  \BibitemOpen
  \bibfield  {author} {\bibinfo {author} {\bibfnamefont {R.~F.}\ \bibnamefont
  {Nalewajski}},\ }\href@noop {} {\emph {\bibinfo {title} {Density {Functional}
  {Theory} {II}: {Relativistic} and {Time} {Dependent} {Extensions}}}}\
  (\bibinfo  {publisher} {Springer Verlag},\ \bibinfo {year}
  {1996})\BibitemShut {NoStop}%
\bibitem [{\citenamefont {Anisimov}\ \emph {et~al.}(1991)\citenamefont
  {Anisimov}, \citenamefont {Zaanen},\ and\ \citenamefont
  {Andersen}}]{anisimov_band_1991}%
  \BibitemOpen
  \bibfield  {author} {\bibinfo {author} {\bibfnamefont {V.~I.}\ \bibnamefont
  {Anisimov}}, \bibinfo {author} {\bibfnamefont {J.}~\bibnamefont {Zaanen}}, \
  and\ \bibinfo {author} {\bibfnamefont {O.~K.}\ \bibnamefont {Andersen}},\
  }\href {\doibase 10.1103/PhysRevB.44.943} {\bibfield  {journal} {\bibinfo
  {journal} {Physical Review B}\ }\textbf {\bibinfo {volume} {44}},\ \bibinfo
  {pages} {943} (\bibinfo {year} {1991})}\BibitemShut {NoStop}%
\bibitem [{\citenamefont {Anisimov}\ \emph {et~al.}(1993)\citenamefont
  {Anisimov}, \citenamefont {Solovyev}, \citenamefont {Korotin}, \citenamefont
  {Czy{\.z}yk},\ and\ \citenamefont
  {Sawatzky}}]{anisimov_density-functional_1993}%
  \BibitemOpen
  \bibfield  {author} {\bibinfo {author} {\bibfnamefont {V.~I.}\ \bibnamefont
  {Anisimov}}, \bibinfo {author} {\bibfnamefont {I.~V.}\ \bibnamefont
  {Solovyev}}, \bibinfo {author} {\bibfnamefont {M.~A.}\ \bibnamefont
  {Korotin}}, \bibinfo {author} {\bibfnamefont {M.~T.}\ \bibnamefont
  {Czy{\.z}yk}}, \ and\ \bibinfo {author} {\bibfnamefont {G.~A.}\ \bibnamefont
  {Sawatzky}},\ }\href {\doibase 10.1103/PhysRevB.48.16929} {\bibfield
  {journal} {\bibinfo  {journal} {Physical Review B}\ }\textbf {\bibinfo
  {volume} {48}},\ \bibinfo {pages} {16929} (\bibinfo {year}
  {1993})}\BibitemShut {NoStop}%
\bibitem [{\citenamefont {Liechtenstein}\ \emph {et~al.}(1995)\citenamefont
  {Liechtenstein}, \citenamefont {Anisimov},\ and\ \citenamefont
  {Zaanen}}]{liechtenstein_density-functional_1995}%
  \BibitemOpen
  \bibfield  {author} {\bibinfo {author} {\bibfnamefont {A.~I.}\ \bibnamefont
  {Liechtenstein}}, \bibinfo {author} {\bibfnamefont {V.~I.}\ \bibnamefont
  {Anisimov}}, \ and\ \bibinfo {author} {\bibfnamefont {J.}~\bibnamefont
  {Zaanen}},\ }\href {\doibase 10.1103/PhysRevB.52.R5467} {\bibfield  {journal}
  {\bibinfo  {journal} {Physical Review B}\ }\textbf {\bibinfo {volume} {52}},\
  \bibinfo {pages} {R5467} (\bibinfo {year} {1995})}\BibitemShut {NoStop}%
\bibitem [{\citenamefont {Gunnarsson}\ \emph {et~al.}(1989)\citenamefont
  {Gunnarsson}, \citenamefont {Andersen}, \citenamefont {Jepsen},\ and\
  \citenamefont {Zaanen}}]{gunnarsson_density-functional_1989}%
  \BibitemOpen
  \bibfield  {author} {\bibinfo {author} {\bibfnamefont {O.}~\bibnamefont
  {Gunnarsson}}, \bibinfo {author} {\bibfnamefont {O.~K.}\ \bibnamefont
  {Andersen}}, \bibinfo {author} {\bibfnamefont {O.}~\bibnamefont {Jepsen}}, \
  and\ \bibinfo {author} {\bibfnamefont {J.}~\bibnamefont {Zaanen}},\ }\href
  {\doibase 10.1103/PhysRevB.39.1708} {\bibfield  {journal} {\bibinfo
  {journal} {Physical Review B}\ }\textbf {\bibinfo {volume} {39}},\ \bibinfo
  {pages} {1708} (\bibinfo {year} {1989})}\BibitemShut {NoStop}%
\bibitem [{\citenamefont {Gunnarsson}(1990)}]{gunnarsson_calculation_1990}%
  \BibitemOpen
  \bibfield  {author} {\bibinfo {author} {\bibfnamefont {O.}~\bibnamefont
  {Gunnarsson}},\ }\href {\doibase 10.1103/PhysRevB.41.514} {\bibfield
  {journal} {\bibinfo  {journal} {Physical Review B}\ }\textbf {\bibinfo
  {volume} {41}},\ \bibinfo {pages} {514} (\bibinfo {year} {1990})}\BibitemShut
  {NoStop}%
\bibitem [{\citenamefont {Springer}\ and\ \citenamefont
  {Aryasetiawan}(1998)}]{springer_frequency-dependent_1998}%
  \BibitemOpen
  \bibfield  {author} {\bibinfo {author} {\bibfnamefont {M.}~\bibnamefont
  {Springer}}\ and\ \bibinfo {author} {\bibfnamefont {F.}~\bibnamefont
  {Aryasetiawan}},\ }\href {\doibase 10.1103/PhysRevB.57.4364} {\bibfield
  {journal} {\bibinfo  {journal} {Physical Review B}\ }\textbf {\bibinfo
  {volume} {57}},\ \bibinfo {pages} {4364} (\bibinfo {year}
  {1998})}\BibitemShut {NoStop}%
\bibitem [{\citenamefont {Kotani}(2000)}]{kotani_ab_2000}%
  \BibitemOpen
  \bibfield  {author} {\bibinfo {author} {\bibfnamefont {T.}~\bibnamefont
  {Kotani}},\ }\href {\doibase 10.1088/0953-8984/12/11/307} {\bibfield
  {journal} {\bibinfo  {journal} {J. Phys.: Condens. Matter}\ }\textbf
  {\bibinfo {volume} {12}},\ \bibinfo {pages} {2413} (\bibinfo {year}
  {2000})}\BibitemShut {NoStop}%
\bibitem [{\citenamefont {Aryasetiawan}\ \emph {et~al.}(2004)\citenamefont
  {Aryasetiawan}, \citenamefont {Imada}, \citenamefont {Georges}, \citenamefont
  {Kotliar}, \citenamefont {Biermann},\ and\ \citenamefont
  {Lichtenstein}}]{aryasetiawan_frequency-dependent_2004}%
  \BibitemOpen
  \bibfield  {author} {\bibinfo {author} {\bibfnamefont {F.}~\bibnamefont
  {Aryasetiawan}}, \bibinfo {author} {\bibfnamefont {M.}~\bibnamefont {Imada}},
  \bibinfo {author} {\bibfnamefont {A.}~\bibnamefont {Georges}}, \bibinfo
  {author} {\bibfnamefont {G.}~\bibnamefont {Kotliar}}, \bibinfo {author}
  {\bibfnamefont {S.}~\bibnamefont {Biermann}}, \ and\ \bibinfo {author}
  {\bibfnamefont {A.~I.}\ \bibnamefont {Lichtenstein}},\ }\href {\doibase
  10.1103/PhysRevB.70.195104} {\bibfield  {journal} {\bibinfo  {journal}
  {Physical Review B}\ }\textbf {\bibinfo {volume} {70}},\ \bibinfo {pages}
  {195104} (\bibinfo {year} {2004})}\BibitemShut {NoStop}%
\bibitem [{\citenamefont {Aryasetiawan}\ \emph {et~al.}(2006)\citenamefont
  {Aryasetiawan}, \citenamefont {Karlsson}, \citenamefont {Jepsen},\ and\
  \citenamefont {Sch{\"o}nberger}}]{aryasetiawan_calculations_2006}%
  \BibitemOpen
  \bibfield  {author} {\bibinfo {author} {\bibfnamefont {F.}~\bibnamefont
  {Aryasetiawan}}, \bibinfo {author} {\bibfnamefont {K.}~\bibnamefont
  {Karlsson}}, \bibinfo {author} {\bibfnamefont {O.}~\bibnamefont {Jepsen}}, \
  and\ \bibinfo {author} {\bibfnamefont {U.}~\bibnamefont {Sch{\"o}nberger}},\
  }\href {\doibase 10.1103/PhysRevB.74.125106} {\bibfield  {journal} {\bibinfo
  {journal} {Physical Review B}\ }\textbf {\bibinfo {volume} {74}},\ \bibinfo
  {pages} {125106} (\bibinfo {year} {2006})}\BibitemShut {NoStop}%
\bibitem [{\citenamefont {Solovyev}\ and\ \citenamefont
  {Imada}(2005)}]{solovyev_screening_2005}%
  \BibitemOpen
  \bibfield  {author} {\bibinfo {author} {\bibfnamefont {I.~V.}\ \bibnamefont
  {Solovyev}}\ and\ \bibinfo {author} {\bibfnamefont {M.}~\bibnamefont
  {Imada}},\ }\href {\doibase 10.1103/PhysRevB.71.045103} {\bibfield  {journal}
  {\bibinfo  {journal} {Physical Review B}\ }\textbf {\bibinfo {volume} {71}},\
  \bibinfo {pages} {045103} (\bibinfo {year} {2005})}\BibitemShut {NoStop}%
\bibitem [{\citenamefont {Skone}\ \emph {et~al.}(2016)\citenamefont {Skone},
  \citenamefont {Govoni},\ and\ \citenamefont
  {Galli}}]{skone_nonempirical_2016}%
  \BibitemOpen
  \bibfield  {author} {\bibinfo {author} {\bibfnamefont {J.~H.}\ \bibnamefont
  {Skone}}, \bibinfo {author} {\bibfnamefont {M.}~\bibnamefont {Govoni}}, \
  and\ \bibinfo {author} {\bibfnamefont {G.}~\bibnamefont {Galli}},\ }\href
  {\doibase 10.1103/PhysRevB.93.235106} {\bibfield  {journal} {\bibinfo
  {journal} {Physical Review B}\ }\textbf {\bibinfo {volume} {93}},\ \bibinfo
  {pages} {235106} (\bibinfo {year} {2016})}\BibitemShut {NoStop}%
\bibitem [{\citenamefont {Brawand}\ \emph {et~al.}(2017)\citenamefont
  {Brawand}, \citenamefont {Govoni}, \citenamefont {V{\"o}r{\"o}s},\ and\
  \citenamefont {Galli}}]{brawand_performance_2017}%
  \BibitemOpen
  \bibfield  {author} {\bibinfo {author} {\bibfnamefont {N.~P.}\ \bibnamefont
  {Brawand}}, \bibinfo {author} {\bibfnamefont {M.}~\bibnamefont {Govoni}},
  \bibinfo {author} {\bibfnamefont {M.}~\bibnamefont {V{\"o}r{\"o}s}}, \ and\
  \bibinfo {author} {\bibfnamefont {G.}~\bibnamefont {Galli}},\ }\href
  {\doibase 10.1021/acs.jctc.7b00368} {\bibfield  {journal} {\bibinfo
  {journal} {J. Chem. Theory Comput.}\ } (\bibinfo {year} {2017}),\
  10.1021/acs.jctc.7b00368}\BibitemShut {NoStop}%
\bibitem [{\citenamefont {K{\"o}rbel}\ \emph {et~al.}(2014)\citenamefont
  {K{\"o}rbel}, \citenamefont {Boulanger}, \citenamefont {Duchemin},
  \citenamefont {Blase}, \citenamefont {Marques},\ and\ \citenamefont
  {Botti}}]{korbel_benchmark_2014}%
  \BibitemOpen
  \bibfield  {author} {\bibinfo {author} {\bibfnamefont {S.}~\bibnamefont
  {K{\"o}rbel}}, \bibinfo {author} {\bibfnamefont {P.}~\bibnamefont
  {Boulanger}}, \bibinfo {author} {\bibfnamefont {I.}~\bibnamefont {Duchemin}},
  \bibinfo {author} {\bibfnamefont {X.}~\bibnamefont {Blase}}, \bibinfo
  {author} {\bibfnamefont {M.~A.~L.}\ \bibnamefont {Marques}}, \ and\ \bibinfo
  {author} {\bibfnamefont {S.}~\bibnamefont {Botti}},\ }\href {\doibase
  10.1021/ct5003658} {\bibfield  {journal} {\bibinfo  {journal} {J. Chem.
  Theory Comput.}\ }\textbf {\bibinfo {volume} {10}},\ \bibinfo {pages} {3934}
  (\bibinfo {year} {2014})}\BibitemShut {NoStop}%
\bibitem [{\citenamefont {Baroni}\ \emph {et~al.}(2001)\citenamefont {Baroni},
  \citenamefont {de~Gironcoli}, \citenamefont {Dal~Corso},\ and\ \citenamefont
  {Giannozzi}}]{baroni_phonons_2001}%
  \BibitemOpen
  \bibfield  {author} {\bibinfo {author} {\bibfnamefont {S.}~\bibnamefont
  {Baroni}}, \bibinfo {author} {\bibfnamefont {S.}~\bibnamefont
  {de~Gironcoli}}, \bibinfo {author} {\bibfnamefont {A.}~\bibnamefont
  {Dal~Corso}}, \ and\ \bibinfo {author} {\bibfnamefont {P.}~\bibnamefont
  {Giannozzi}},\ }\href {\doibase 10.1103/RevModPhys.73.515} {\bibfield
  {journal} {\bibinfo  {journal} {Reviews of Modern Physics}\ }\textbf
  {\bibinfo {volume} {73}},\ \bibinfo {pages} {515} (\bibinfo {year}
  {2001})}\BibitemShut {NoStop}%
\bibitem [{\citenamefont {Giannozzi}\ \emph {et~al.}(2009)\citenamefont
  {Giannozzi}, \citenamefont {Baroni}, \citenamefont {Bonini}, \citenamefont
  {Calandra}, \citenamefont {Car}, \citenamefont {Cavazzoni}, \citenamefont
  {Ceresoli}, \citenamefont {Chiarotti}, \citenamefont {Cococcioni},
  \citenamefont {Dabo}, \citenamefont {Corso}, \citenamefont {Gironcoli},
  \citenamefont {Fabris}, \citenamefont {Fratesi}, \citenamefont {Gebauer},
  \citenamefont {Gerstmann}, \citenamefont {Gougoussis}, \citenamefont
  {Kokalj}, \citenamefont {Lazzeri}, \citenamefont {Martin-Samos},
  \citenamefont {Marzari}, \citenamefont {Mauri}, \citenamefont {Mazzarello},
  \citenamefont {Paolini}, \citenamefont {Pasquarello}, \citenamefont
  {Paulatto}, \citenamefont {Sbraccia}, \citenamefont {Scandolo}, \citenamefont
  {Sclauzero}, \citenamefont {Seitsonen}, \citenamefont {Smogunov},
  \citenamefont {Umari},\ and\ \citenamefont
  {Wentzcovitch}}]{giannozzi_quantum_2009}%
  \BibitemOpen
  \bibfield  {author} {\bibinfo {author} {\bibfnamefont {P.}~\bibnamefont
  {Giannozzi}}, \bibinfo {author} {\bibfnamefont {S.}~\bibnamefont {Baroni}},
  \bibinfo {author} {\bibfnamefont {N.}~\bibnamefont {Bonini}}, \bibinfo
  {author} {\bibfnamefont {M.}~\bibnamefont {Calandra}}, \bibinfo {author}
  {\bibfnamefont {R.}~\bibnamefont {Car}}, \bibinfo {author} {\bibfnamefont
  {C.}~\bibnamefont {Cavazzoni}}, \bibinfo {author} {\bibfnamefont
  {D.}~\bibnamefont {Ceresoli}}, \bibinfo {author} {\bibfnamefont {G.~L.}\
  \bibnamefont {Chiarotti}}, \bibinfo {author} {\bibfnamefont {M.}~\bibnamefont
  {Cococcioni}}, \bibinfo {author} {\bibfnamefont {I.}~\bibnamefont {Dabo}},
  \bibinfo {author} {\bibfnamefont {A.~D.}\ \bibnamefont {Corso}}, \bibinfo
  {author} {\bibfnamefont {S.~d.}\ \bibnamefont {Gironcoli}}, \bibinfo {author}
  {\bibfnamefont {S.}~\bibnamefont {Fabris}}, \bibinfo {author} {\bibfnamefont
  {G.}~\bibnamefont {Fratesi}}, \bibinfo {author} {\bibfnamefont
  {R.}~\bibnamefont {Gebauer}}, \bibinfo {author} {\bibfnamefont
  {U.}~\bibnamefont {Gerstmann}}, \bibinfo {author} {\bibfnamefont
  {C.}~\bibnamefont {Gougoussis}}, \bibinfo {author} {\bibfnamefont
  {A.}~\bibnamefont {Kokalj}}, \bibinfo {author} {\bibfnamefont
  {M.}~\bibnamefont {Lazzeri}}, \bibinfo {author} {\bibfnamefont
  {L.}~\bibnamefont {Martin-Samos}}, \bibinfo {author} {\bibfnamefont
  {N.}~\bibnamefont {Marzari}}, \bibinfo {author} {\bibfnamefont
  {F.}~\bibnamefont {Mauri}}, \bibinfo {author} {\bibfnamefont
  {R.}~\bibnamefont {Mazzarello}}, \bibinfo {author} {\bibfnamefont
  {S.}~\bibnamefont {Paolini}}, \bibinfo {author} {\bibfnamefont
  {A.}~\bibnamefont {Pasquarello}}, \bibinfo {author} {\bibfnamefont
  {L.}~\bibnamefont {Paulatto}}, \bibinfo {author} {\bibfnamefont
  {C.}~\bibnamefont {Sbraccia}}, \bibinfo {author} {\bibfnamefont
  {S.}~\bibnamefont {Scandolo}}, \bibinfo {author} {\bibfnamefont
  {G.}~\bibnamefont {Sclauzero}}, \bibinfo {author} {\bibfnamefont {A.~P.}\
  \bibnamefont {Seitsonen}}, \bibinfo {author} {\bibfnamefont {A.}~\bibnamefont
  {Smogunov}}, \bibinfo {author} {\bibfnamefont {P.}~\bibnamefont {Umari}}, \
  and\ \bibinfo {author} {\bibfnamefont {R.~M.}\ \bibnamefont {Wentzcovitch}},\
  }\href {\doibase 10.1088/0953-8984/21/39/395502} {\bibfield  {journal}
  {\bibinfo  {journal} {Journal of Physics: Condensed Matter}\ }\textbf
  {\bibinfo {volume} {21}},\ \bibinfo {pages} {395502} (\bibinfo {year}
  {2009})}\BibitemShut {NoStop}%
\bibitem [{\citenamefont {Giannozzi}\ \emph {et~al.}(2017)\citenamefont
  {Giannozzi}, \citenamefont {Andreussi}, \citenamefont {Brumme}, \citenamefont
  {Bunau}, \citenamefont {Nardelli}, \citenamefont {Calandra}, \citenamefont
  {Car}, \citenamefont {Cavazzoni}, \citenamefont {{D Ceresoli}}, \citenamefont
  {Cococcioni}, \citenamefont {Colonna}, \citenamefont {Carnimeo},
  \citenamefont {Corso}, \citenamefont {Gironcoli}, \citenamefont {Delugas},
  \citenamefont {Jr}, \citenamefont {{A Ferretti}}, \citenamefont {Floris},
  \citenamefont {Fratesi}, \citenamefont {Fugallo}, \citenamefont {Gebauer},
  \citenamefont {Gerstmann}, \citenamefont {Giustino}, \citenamefont {Gorni},
  \citenamefont {Jia}, \citenamefont {Kawamura}, \citenamefont {{H-Y Ko}},
  \citenamefont {Kokalj}, \citenamefont {K{\"u}{\c c}{\"u}kbenli},
  \citenamefont {Lazzeri}, \citenamefont {Marsili}, \citenamefont {Marzari},
  \citenamefont {Mauri}, \citenamefont {Nguyen}, \citenamefont {Nguyen},
  \citenamefont {{A Otero-de-la-Roza}}, \citenamefont {Paulatto}, \citenamefont
  {Ponc{\'e}}, \citenamefont {Rocca}, \citenamefont {Sabatini}, \citenamefont
  {Santra}, \citenamefont {Schlipf}, \citenamefont {Seitsonen}, \citenamefont
  {Smogunov}, \citenamefont {{I Timrov}}, \citenamefont {Thonhauser},
  \citenamefont {Umari}, \citenamefont {Vast}, \citenamefont {Wu},\ and\
  \citenamefont {Baroni}}]{giannozzi_advanced_2017}%
  \BibitemOpen
  \bibfield  {author} {\bibinfo {author} {\bibfnamefont {P.}~\bibnamefont
  {Giannozzi}}, \bibinfo {author} {\bibfnamefont {O.}~\bibnamefont
  {Andreussi}}, \bibinfo {author} {\bibfnamefont {T.}~\bibnamefont {Brumme}},
  \bibinfo {author} {\bibfnamefont {O.}~\bibnamefont {Bunau}}, \bibinfo
  {author} {\bibfnamefont {M.~B.}\ \bibnamefont {Nardelli}}, \bibinfo {author}
  {\bibfnamefont {M.}~\bibnamefont {Calandra}}, \bibinfo {author}
  {\bibfnamefont {R.}~\bibnamefont {Car}}, \bibinfo {author} {\bibfnamefont
  {C.}~\bibnamefont {Cavazzoni}}, \bibinfo {author} {\bibnamefont {{D
  Ceresoli}}}, \bibinfo {author} {\bibfnamefont {M.}~\bibnamefont
  {Cococcioni}}, \bibinfo {author} {\bibfnamefont {N.}~\bibnamefont {Colonna}},
  \bibinfo {author} {\bibfnamefont {I.}~\bibnamefont {Carnimeo}}, \bibinfo
  {author} {\bibfnamefont {A.~D.}\ \bibnamefont {Corso}}, \bibinfo {author}
  {\bibfnamefont {S.~d.}\ \bibnamefont {Gironcoli}}, \bibinfo {author}
  {\bibfnamefont {P.}~\bibnamefont {Delugas}}, \bibinfo {author} {\bibfnamefont
  {R.~A.~D.}\ \bibnamefont {Jr}}, \bibinfo {author} {\bibnamefont {{A
  Ferretti}}}, \bibinfo {author} {\bibfnamefont {A.}~\bibnamefont {Floris}},
  \bibinfo {author} {\bibfnamefont {G.}~\bibnamefont {Fratesi}}, \bibinfo
  {author} {\bibfnamefont {G.}~\bibnamefont {Fugallo}}, \bibinfo {author}
  {\bibfnamefont {R.}~\bibnamefont {Gebauer}}, \bibinfo {author} {\bibfnamefont
  {U.}~\bibnamefont {Gerstmann}}, \bibinfo {author} {\bibfnamefont
  {F.}~\bibnamefont {Giustino}}, \bibinfo {author} {\bibfnamefont
  {T.}~\bibnamefont {Gorni}}, \bibinfo {author} {\bibfnamefont
  {J.}~\bibnamefont {Jia}}, \bibinfo {author} {\bibfnamefont {M.}~\bibnamefont
  {Kawamura}}, \bibinfo {author} {\bibnamefont {{H-Y Ko}}}, \bibinfo {author}
  {\bibfnamefont {A.}~\bibnamefont {Kokalj}}, \bibinfo {author} {\bibfnamefont
  {E.}~\bibnamefont {K{\"u}{\c c}{\"u}kbenli}}, \bibinfo {author}
  {\bibfnamefont {M.}~\bibnamefont {Lazzeri}}, \bibinfo {author} {\bibfnamefont
  {M.}~\bibnamefont {Marsili}}, \bibinfo {author} {\bibfnamefont
  {N.}~\bibnamefont {Marzari}}, \bibinfo {author} {\bibfnamefont
  {F.}~\bibnamefont {Mauri}}, \bibinfo {author} {\bibfnamefont {N.~L.}\
  \bibnamefont {Nguyen}}, \bibinfo {author} {\bibfnamefont {H.-V.}\
  \bibnamefont {Nguyen}}, \bibinfo {author} {\bibnamefont {{A
  Otero-de-la-Roza}}}, \bibinfo {author} {\bibfnamefont {L.}~\bibnamefont
  {Paulatto}}, \bibinfo {author} {\bibfnamefont {S.}~\bibnamefont {Ponc{\'e}}},
  \bibinfo {author} {\bibfnamefont {D.}~\bibnamefont {Rocca}}, \bibinfo
  {author} {\bibfnamefont {R.}~\bibnamefont {Sabatini}}, \bibinfo {author}
  {\bibfnamefont {B.}~\bibnamefont {Santra}}, \bibinfo {author} {\bibfnamefont
  {M.}~\bibnamefont {Schlipf}}, \bibinfo {author} {\bibfnamefont {A.~P.}\
  \bibnamefont {Seitsonen}}, \bibinfo {author} {\bibfnamefont {A.}~\bibnamefont
  {Smogunov}}, \bibinfo {author} {\bibnamefont {{I Timrov}}}, \bibinfo {author}
  {\bibfnamefont {T.}~\bibnamefont {Thonhauser}}, \bibinfo {author}
  {\bibfnamefont {P.}~\bibnamefont {Umari}}, \bibinfo {author} {\bibfnamefont
  {N.}~\bibnamefont {Vast}}, \bibinfo {author} {\bibfnamefont {X.}~\bibnamefont
  {Wu}}, \ and\ \bibinfo {author} {\bibfnamefont {S.}~\bibnamefont {Baroni}},\
  }\href {\doibase 10.1088/1361-648X/aa8f79} {\bibfield  {journal} {\bibinfo
  {journal} {Journal of Physics: Condensed Matter}\ }\textbf {\bibinfo {volume}
  {29}},\ \bibinfo {pages} {465901} (\bibinfo {year} {2017})}\BibitemShut
  {NoStop}%
\bibitem [{\citenamefont {Hamann}(2013)}]{hamann_optimized_2013}%
  \BibitemOpen
  \bibfield  {author} {\bibinfo {author} {\bibfnamefont {D.~R.}\ \bibnamefont
  {Hamann}},\ }\href {\doibase 10.1103/PhysRevB.88.085117} {\bibfield
  {journal} {\bibinfo  {journal} {Physical Review B}\ }\textbf {\bibinfo
  {volume} {88}},\ \bibinfo {pages} {085117} (\bibinfo {year}
  {2013})}\BibitemShut {NoStop}%
\bibitem [{\citenamefont {Schlipf}\ and\ \citenamefont
  {Gygi}(2015)}]{schlipf_optimization_2015}%
  \BibitemOpen
  \bibfield  {author} {\bibinfo {author} {\bibfnamefont {M.}~\bibnamefont
  {Schlipf}}\ and\ \bibinfo {author} {\bibfnamefont {F.}~\bibnamefont {Gygi}},\
  }\href {\doibase 10.1016/j.cpc.2015.05.011} {\bibfield  {journal} {\bibinfo
  {journal} {Computer Physics Communications}\ }\textbf {\bibinfo {volume}
  {196}},\ \bibinfo {pages} {36} (\bibinfo {year} {2015})}\BibitemShut
  {NoStop}%
\bibitem [{ONC(2017)}]{ONCV_website}%
  \BibitemOpen
  \href {http://www.quantum-simulation.org/potentials/sg15_oncv/} {\enquote
  {\bibinfo {title} {{SG15} {ONCV} {Potentials}},}\ } (\bibinfo {year}
  {2017})\BibitemShut {NoStop}%
\bibitem [{Note2()}]{Note2}%
  \BibitemOpen
  \bibinfo {note} {The equilibrium geometries of the transition metal complexes
  were kindly provided by the authors of Ref.~\protect \citenum
  {korbel_benchmark_2014}.}\BibitemShut {Stop}%
\bibitem [{\citenamefont {Li}\ and\ \citenamefont
  {Dabo}(2011)}]{li_electronic_2011}%
  \BibitemOpen
  \bibfield  {author} {\bibinfo {author} {\bibfnamefont {Y.}~\bibnamefont
  {Li}}\ and\ \bibinfo {author} {\bibfnamefont {I.}~\bibnamefont {Dabo}},\
  }\href {\doibase 10.1103/PhysRevB.84.155127} {\bibfield  {journal} {\bibinfo
  {journal} {Physical Review B}\ }\textbf {\bibinfo {volume} {84}},\ \bibinfo
  {pages} {155127} (\bibinfo {year} {2011})}\BibitemShut {NoStop}%
\bibitem [{\citenamefont {Perdew}\ \emph {et~al.}(1996)\citenamefont {Perdew},
  \citenamefont {Burke},\ and\ \citenamefont
  {Ernzerhof}}]{perdew_generalized_1996}%
  \BibitemOpen
  \bibfield  {author} {\bibinfo {author} {\bibfnamefont {J.~P.}\ \bibnamefont
  {Perdew}}, \bibinfo {author} {\bibfnamefont {K.}~\bibnamefont {Burke}}, \
  and\ \bibinfo {author} {\bibfnamefont {M.}~\bibnamefont {Ernzerhof}},\ }\href
  {\doibase 10.1103/PhysRevLetters77.3865} {\bibfield  {journal} {\bibinfo
  {journal} {Physical Review Letters}\ }\textbf {\bibinfo {volume} {77}},\
  \bibinfo {pages} {3865} (\bibinfo {year} {1996})}\BibitemShut {NoStop}%
\bibitem [{Note3()}]{Note3}%
  \BibitemOpen
  \bibinfo {note} {For those experiments for which is unclear whether they are
  vertical or adiabatic (see Supporting Information for the complete list) the
  calculated value are left out of the average. In the case of adiabatic
  experimental value we used the same correction used in Ref.~\protect \citenum
  {korbel_benchmark_2014} based on the difference between adiabatic and
  vertical $\Delta $SCF energies.}\BibitemShut {Stop}%
\bibitem [{\citenamefont {Wu}\ \emph {et~al.}(2011)\citenamefont {Wu},
  \citenamefont {Xie}, \citenamefont {Qin}, \citenamefont {Tan}, \citenamefont
  {Tang},\ and\ \citenamefont {Lu}}]{wu_photoelectron_2011}%
  \BibitemOpen
  \bibfield  {author} {\bibinfo {author} {\bibfnamefont {X.}~\bibnamefont
  {Wu}}, \bibinfo {author} {\bibfnamefont {H.}~\bibnamefont {Xie}}, \bibinfo
  {author} {\bibfnamefont {Z.}~\bibnamefont {Qin}}, \bibinfo {author}
  {\bibfnamefont {K.}~\bibnamefont {Tan}}, \bibinfo {author} {\bibfnamefont
  {Z.}~\bibnamefont {Tang}}, \ and\ \bibinfo {author} {\bibfnamefont
  {X.}~\bibnamefont {Lu}},\ }\href {\doibase 10.1021/jp1100686} {\bibfield
  {journal} {\bibinfo  {journal} {J. Phys. Chem. A}\ }\textbf {\bibinfo
  {volume} {115}},\ \bibinfo {pages} {6321} (\bibinfo {year}
  {2011})}\BibitemShut {NoStop}%
\bibitem [{\citenamefont {Linstrom}\ and\ \citenamefont
  {Mallard}(2005)}]{nist2005}%
  \BibitemOpen
  \bibinfo {editor} {\bibfnamefont {P.~J.}\ \bibnamefont {Linstrom}}\ and\
  \bibinfo {editor} {\bibfnamefont {W.~G.}\ \bibnamefont {Mallard}},\ eds.,\
  \href {http://webbook.nist.gov} {\emph {\bibinfo {title} {{NIST Chemistry
  WebBook, NIST Standard Reference Database Number 69}}}}\ (\bibinfo
  {publisher} {National Institute of Standards and Technology},\ \bibinfo
  {address} {Gaithersburg MD, 20899},\ \bibinfo {year} {2005})\BibitemShut
  {NoStop}%
\bibitem [{\citenamefont {Lehtola}\ \emph {et~al.}(2016)\citenamefont
  {Lehtola}, \citenamefont {Head-Gordon},\ and\ \citenamefont
  {J{\'o}nsson}}]{lehtola_complex_2016}%
  \BibitemOpen
  \bibfield  {author} {\bibinfo {author} {\bibfnamefont {S.}~\bibnamefont
  {Lehtola}}, \bibinfo {author} {\bibfnamefont {M.}~\bibnamefont
  {Head-Gordon}}, \ and\ \bibinfo {author} {\bibfnamefont {H.}~\bibnamefont
  {J{\'o}nsson}},\ }\href {\doibase 10.1021/acs.jctc.6b00347} {\bibfield
  {journal} {\bibinfo  {journal} {J. Chem. Theory Comput.}\ }\textbf {\bibinfo
  {volume} {12}},\ \bibinfo {pages} {3195} (\bibinfo {year}
  {2016})}\BibitemShut {NoStop}%
\bibitem [{\citenamefont {Kulik}\ and\ \citenamefont
  {Marzari}(2011)}]{kulik_accurate_2011}%
  \BibitemOpen
  \bibfield  {author} {\bibinfo {author} {\bibfnamefont {H.~J.}\ \bibnamefont
  {Kulik}}\ and\ \bibinfo {author} {\bibfnamefont {N.}~\bibnamefont
  {Marzari}},\ }\href {\doibase 10.1063/1.3660353} {\bibfield  {journal}
  {\bibinfo  {journal} {The Journal of Chemical Physics}\ }\textbf {\bibinfo
  {volume} {135}},\ \bibinfo {pages} {194105} (\bibinfo {year}
  {2011})}\BibitemShut {NoStop}%
\bibitem [{\citenamefont {Hsu}\ \emph {et~al.}(2009)\citenamefont {Hsu},
  \citenamefont {Umemoto}, \citenamefont {Cococcioni},\ and\ \citenamefont
  {Wentzcovitch}}]{hsu_first-principles_2009}%
  \BibitemOpen
  \bibfield  {author} {\bibinfo {author} {\bibfnamefont {H.}~\bibnamefont
  {Hsu}}, \bibinfo {author} {\bibfnamefont {K.}~\bibnamefont {Umemoto}},
  \bibinfo {author} {\bibfnamefont {M.}~\bibnamefont {Cococcioni}}, \ and\
  \bibinfo {author} {\bibfnamefont {R.}~\bibnamefont {Wentzcovitch}},\ }\href
  {\doibase 10.1103/PhysRevB.79.125124} {\bibfield  {journal} {\bibinfo
  {journal} {Physical Review B}\ }\textbf {\bibinfo {volume} {79}},\ \bibinfo
  {pages} {125124} (\bibinfo {year} {2009})}\BibitemShut {NoStop}%
\end{thebibliography}
%

\end{document}